\documentclass[twocolumn]{openjournal}
\usepackage{xcolor}
\usepackage{textgreek}
\usepackage[utf8]{inputenc}
\usepackage[T1]{fontenc}
\usepackage[english]{babel}
\usepackage{booktabs}
\usepackage{hyperref}
\hypersetup{
    unicode, 
    colorlinks=true,
    linkcolor=linkcolor,
    citecolor=linkcolor,
    filecolor=linkcolor,
    urlcolor=linkcolor,
}
\usepackage{color,colortbl}
\definecolor{linkcolor}{rgb}{0.0,0.3,0.5}
\usepackage{tensind}
\tensordelimiter{?}
\DeclareGraphicsExtensions{.bmp,.png,.jpg,.pdf}
\usepackage{verbatim}
\usepackage[normalem]{ulem}
\usepackage{orcidlink}
\usepackage{soul}
\usepackage{float}
\usepackage{multirow}
\usepackage{balance}
\graphicspath{ {./figs/} }

\begin{document}
\title{Characterising the response of an International LOFAR Station}
\author{Letizia Vincetti\orcidlink{0009-0000-5589-0926}}\thanks{\email{vincettl@tcd.ie}}
\affiliation{School of Physics, Trinity College Dublin, College Green, Dublin 2, Ireland}
\affiliation{Astronomy \& Astrophysics Section, Dublin Institute for Advanced Studies, DIAS Dunsink Observatory, Dublin, D15 XR2R, Ireland}
\affiliation{Armagh Observatory \& Planetarium, College Hill, Armagh, BT61 9DG, Northern Ireland, UK}
\author{S.~C.~Susarla\orcidlink{0000-0003-4332-8201}}
\affiliation{Centre for Astronomy, School of Natural Sciences, University of Galway, University Road, Galway, H91 TK33, Ireland}
\author{Owen A. Johnson\orcidlink{0000-0002-5927-0481}}
\affiliation{School of Physics, Trinity College Dublin, College Green, Dublin 2, Ireland}
\affiliation{Breakthrough Listen, University of California, Berkeley, 501 Campbell Hall 3411, Berkeley, CA 94720, USA}
\author{Evan F. Keane\orcidlink{0000-0002-4553-655X}}
\affiliation{School of Physics, Trinity College Dublin, College Green, Dublin 2, Ireland}
\author{David J. McKenna\orcidlink{0000-0001-7185-1310}}
\affiliation{ASTRON, The Netherlands Institute for Radio Astronomy, Oude Hoogeveensedijk 4, 7991 PD Dwingeloo, The Netherlands}
\author{T.~D.~Carozzi\orcidlink{0000-0002-4963-179X}}
\affiliation{Onsala Space Observatory, Chalmers University, Sweden}
\author{Gavin Ramsay} 
\affiliation{Armagh Observatory \& Planetarium, College Hill, Armagh, BT61 9DG, Northern Ireland, UK}
\author{Ois\'{i}n Creaner}
\affiliation{School of Physical Sciences, Dublin City University, Glasnevin Campus, Dublin, D09 K2WA, Ireland}
\author{Peter T. Gallagher\orcidlink{0000-0001-9745-0400}}
\affiliation{Astronomy \& Astrophysics Section, Dublin Institute for Advanced Studies, DIAS Dunsink Observatory, Dublin, D15 XR2R, Ireland}
\affiliation{School of Physics, Trinity College Dublin, College Green, Dublin 2, Ireland}
\author{Joe McCauley\orcidlink{0000-0003-4399-2233}}
\affiliation{School of Physics, Trinity College Dublin, College Green, Dublin 2, Ireland}

\begin{abstract}
    Phased-array radio interferometers with fixed antennas are a highly scalable design which can achieve a large gain. However they are complex systems and challenging to calibrate. Here we examine the response of an International LOFAR Station using long-track observations of an array of flux-stable pulsars. The analyses are relevant for other arrays like the under-construction SKA-Low. In this paper, we investigate the performance of the high-band antennas of the Irish LOFAR station using long-track observations of bright pulsars. In modelling our measured responses, we account for projection effects, the frequency-dependence of the aperture efficiency and pulsar spectra, as well as sky and instrumental noise contributions. We perform full-track observations of 11 pulsars as they move across the sky. We perform RFI mitigation and determine the signal-to-noise ratio response as a function of time and evaluate the elevation and azimuth dependence. We use the \texttt{DreamBeam} software to model the beam response of the station to compare with what is observed. The sensitivity map so obtained was validated using observations of PSR~B0329+54 on the Swedish LOFAR station. We observe the best response for pointings close to the zenith, as expected, implying an improved sensitivity at higher elevation. However, an asymmetry with respect to the zenith point is detected. Characterising the instrumental response in azimuth reveals a better performance as the targets rise, as compared to when they set. This is the case for all pulsars in the sample used for the Irish station and is also seen with the Swedish station; it also agrees with previously reported results from the Ba\l dy LOFAR station in Poland. The magnitude of the effect can exceed $20\%$. We consider variations in the beam model and noise level, both frequency- and time- dependent, to try to account for this. A possible explanation for the hysteresis-like response in elevation could be an imbalanced signal weighting of the polarisation components. This work highlights the variation in altitude and azimuth of the response of an international LOFAR station using pulsar observations, highlighting an asymmetric response. This work addresses the importance of accurate beam modelling and noise evaluation for the LOFAR 2.0 system upgrade as well as being relevant to the under-construction SKA-Low facility.
\end{abstract}

% Write your keywords here
\begin{keywords}
    {radio astronomy, data analysis, pulsars}
\end{keywords}

\maketitle

\section{Introduction}
The LOw Frequency ARray~\citep[LOFAR,][]{Van_Haarlem_2013} is a hierarchical radio telescope array with frequency coverage of $10$ to $240$~MHz; the antennas are fixed on the ground and grouped into sub-arrays termed `stations', which are distributed across Europe. At any one time, one of two different instruments can be used, depending on the desired frequency range. The Low Band Antennas (LBA) are droopy dipoles and operate at frequencies from $10$ to $90$~MHz, where the antenna elements consist of two inverted V-shaped dipoles. The High Band Antenna (HBA) tiles are used at higher frequencies ranging from $110$ to $240$~MHz. Each tile consists of $16$ HBAs, where each of these consists of two orthogonal bow-tie dipoles. The response of a LOFAR station is complex and depends on a number of factors, such as the effective collecting area, the beam-forming algorithm, contamination due to radio frequency interference (RFI) and instrumental noise \citep[e.g.][]{Kondratiev_2016}. Additional complexity arises from the much larger field of view (FoV) at LOFAR frequencies, which exceeds the scale of spatial variations in the sky temperature and typically encompasses bright sources of astrophysical and/or terrestrial origin \citep{Carozzi_2009}. 

LOFAR can operate stations across Europe as a wide interferometric array ~\citep{Shimwell_2019}, but also in single station stand-alone mode. With this paper, we address the lack of a well characterized response of an international LOFAR station.

Despite its powerful capabilities and large FoV, substantial effort has gone into developing a calibration procedure to correct distortions in the received signal (e.g. ~\citet{Brackenhoff_2025, deGasperin_2019, Corstanje_2016, Kondratiev_2016}). In particular, observations below an elevation of $\sim30^{\circ}$ are highly discouraged, as the sensitivity drops significantly due to the reduced gain of the dipoles and reduced phased-array effective collecting area projected on the sky ~\citep{Van_Haarlem_2013}. Given the elevation-dependent sensitivity and the complexity of the beam modelling and calibration, understanding the station response would be advantageous for a number of reasons. First of all, improved beam modelling, together with direction-dependent and direction-independent antenna gains, and reliable calibration solutions would lead inevitably to accurate flux density measurements and reduced polarisation leakage (see e.g. \citet{Asad_2015}). Secondly, from an operational point of view, the evaluation of the response of a LOFAR station would enable more effective observation scheduling, hence an efficient use of the telescope's resources and increased scientific output. 

In this paper we use the Irish LOFAR station to characterize the response of the HBAs using long-track observations of $11$ bright pulsars, for a total of $170$~observing hours. We model our observations using the Radio Interferometer Measurement Equation (RIME) formalism as in \citet{SmirnovI_2011,Carozzi_2016}. Pulsars are rapidly rotating and strongly magnetized neutron stars \citep{Gold_1968}. Powered by their rotational energy, their emission in the radio band is detected in the form of pulses \citep{Kramer_2003}. Most pulsars are seen to have stable flux densities~\citep{Jankowski_2018}; we use a sample at sufficient distance to be unaffected by propagation effects like interstellar scintillation ~\citep{Wu_2022}. In this work the pulses are coherently folded on top of one another to form a stable high signal-to-noise ratio time-integrated pulse profile ~\citep{Handbook_2004}. 

In Sect.~\ref{sec:ilofar}, we describe the contributing factors to the instrumental response of a LOFAR HBA station and describe the parameters of the Irish LOFAR station. In Sect.~\ref{sec:data} we explain the observing targets and describe the data acquisition and subsequent processing methods. We compare our results with those obtained using the Polish station (PL612) in ~\citet{Blaszkiewicz_2018} for two pulsars (PSRs~B0329$+$54 and B1508$+$55) observed from both sites. Observations of PSR~B0329$+$54 were also taken with the HBA station at the Swedish LOFAR site (SE607) and are presented alongside the Irish station results. In Sect.~\ref{sec:results} we present the results of our analyses and methods at determining a self-consistent station response. We then highlight the key results of this study in Sect.~\ref{sec:discussion}, along with operational recommendations (Sect.~\ref{conclusions}) for calbrating SKA-Low~\citep{Braun2015ska}, which is currently under construction. 
\begin{center}
\begin{figure*}[t!]
   \includegraphics[width=6cm]{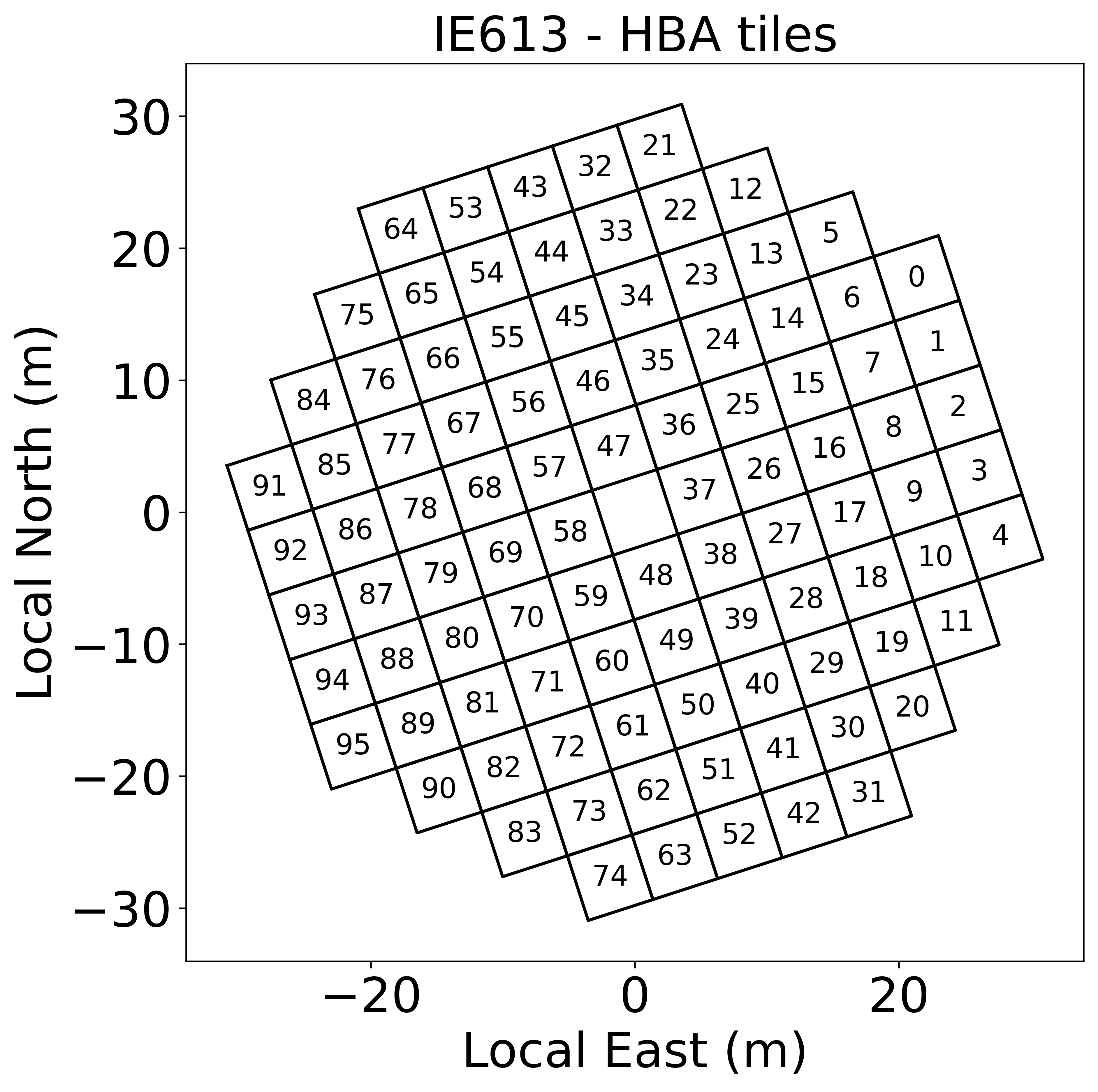}
   \includegraphics[width=6cm]{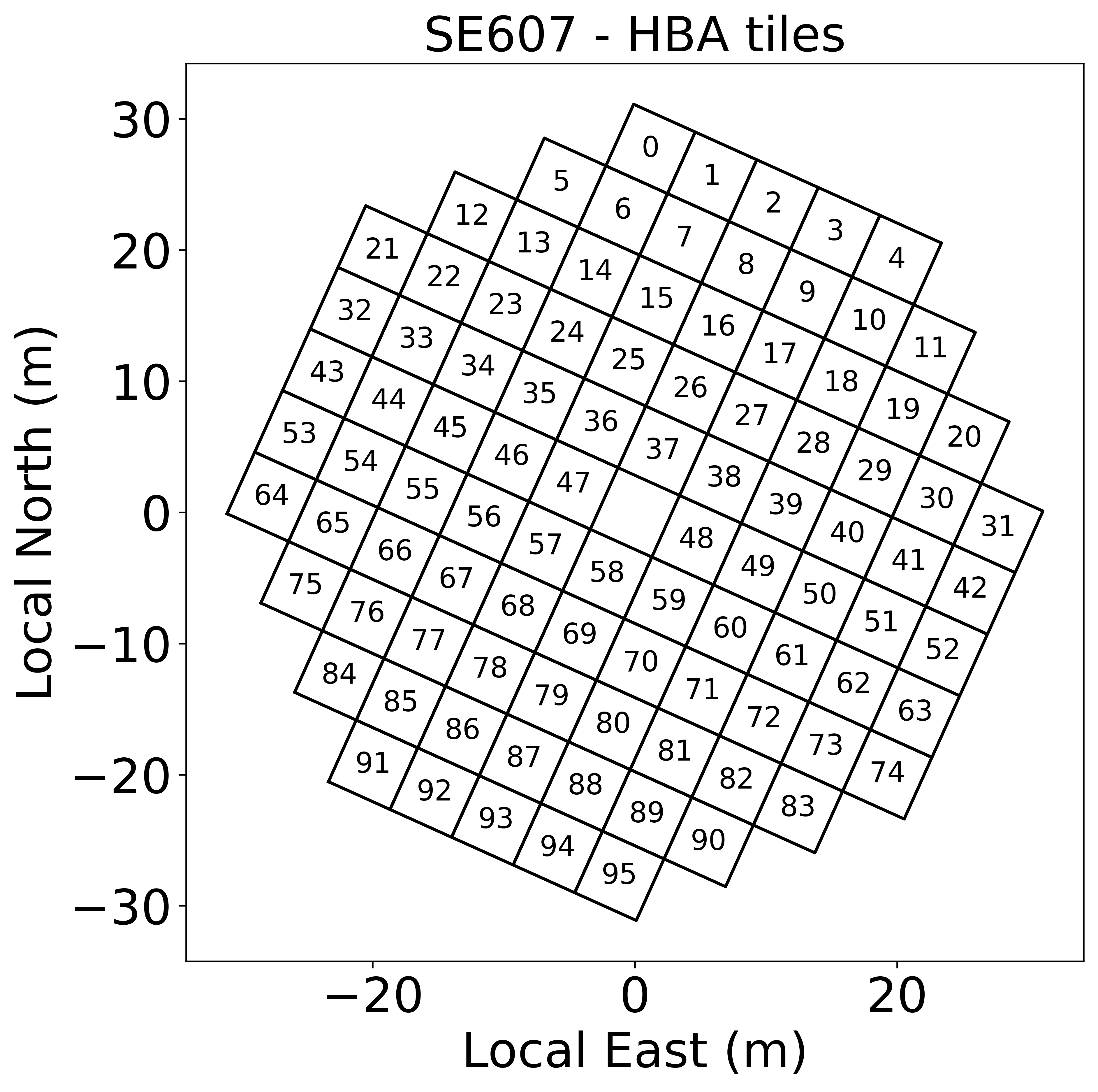}
   \includegraphics[width=6cm]{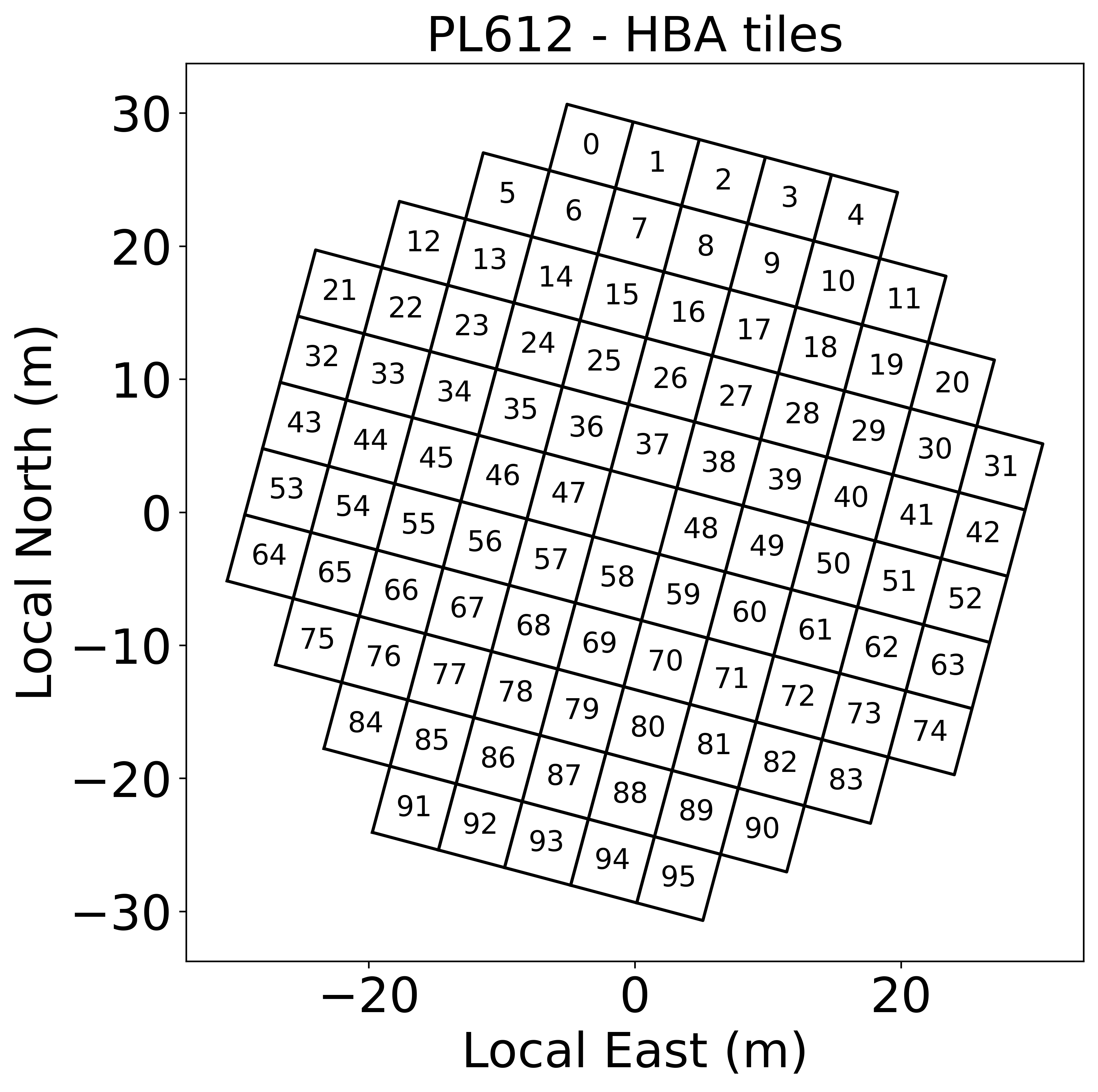}
\caption{From left to right: the layout of IE613, SE607, PL612 HBAs (the rotation of the array is $11.90^{\circ}$ from local north (IE613), 4.26$^{\circ}$ (SE607) and 10.98$^{\circ}$(PL612).}
\label{layout}
\end{figure*}
\end{center}
\section{The Irish LOFAR station}\label{sec:ilofar}

The Irish LOFAR station (I-LOFAR; hereafter IE613) is one of the $14$ international stations of the LOFAR telescope, located at the Rosse Observatory in Birr, Co. Offaly, Ireland. It is made up of $96$ dual-polarisation LBAs operating in the frequency range $10-90$~MHz and $96$ HBA tiles operating in the range $110-240$ MHz.

The HBA array is the focus of this work, its orientation on the ground is shown in the left-hand panel of Fig. \ref{layout}. Made of 96 tiles, $5\mathrm{m}\times5$m in size, for a total of $1536\times2$ bow-tie antennas, the array has a diameter of $56.5$m. \\
The HBA layout is designed specifically to increase the gain of a single antenna and consequently to maximize the effective collecting area of the station, up to $2400$~$\rm m^2$ at $120$~MHz. The dipoles are arranged close enough to suppress grating lobes, but spaced enough to allow for a sufficiently narrow beam for the entire station. For the purpose of the work that follows, we next summarize how the signal is acquired and processed by the IE613 HBA station (Sect.~\ref{station_bf}) and how the single station sensitivity is derived (Sect.~\ref{station_sens}).

\subsection{Station Beamforming}\label{station_bf}
\begin{figure}[h]
    \centering
    \includegraphics[width=\hsize]{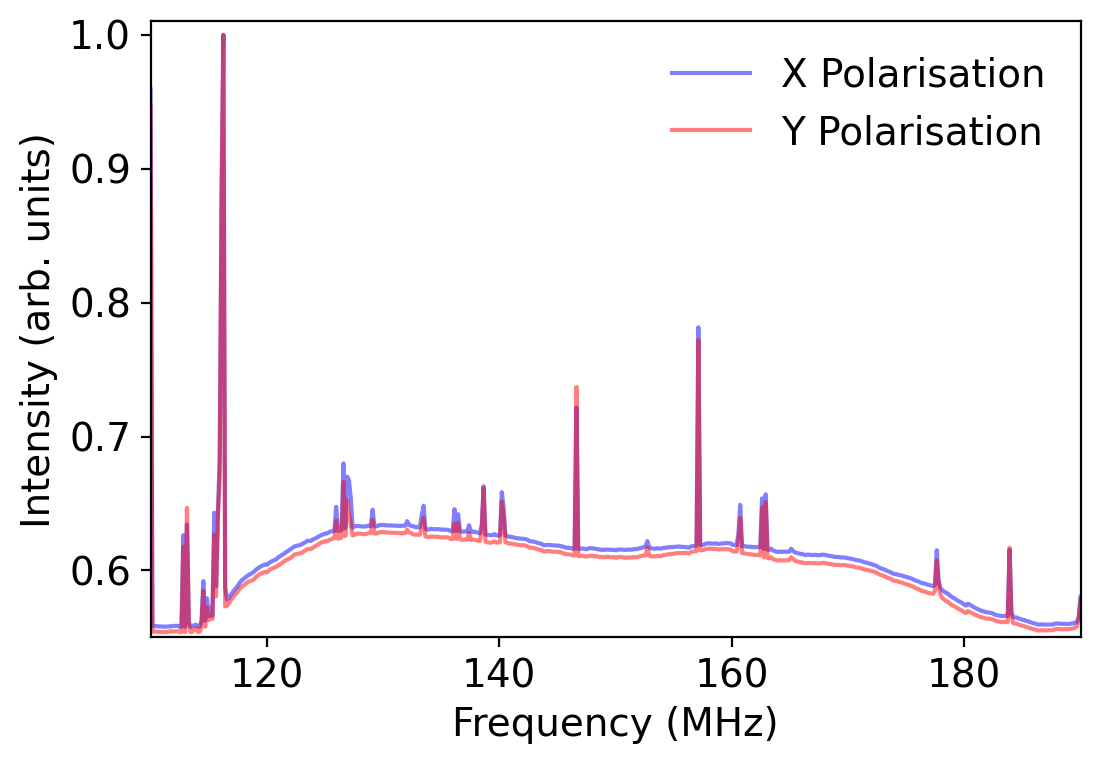}
    \caption{Example of antenna spectra generated as part of the diagnostics and monitoring procedures each time an observation is scheduled. The plot is generated with 1s integration of data and varies with date and time ($06:01$ UTC on $14$ February $2025$ the one shown here). The X and Y polarisation responses shown are the averaged ones over all antennas.}
    \label{fig:bandpass}
\end{figure}

Several different operating modes are available for observations with a LOFAR telescope. The beam-formed mode allows for the recording of a time-series dynamic spectra data product. The beamforming refers to the coherent addition of antenna signals with appropriate time and phase delays in order to point at a
specific location in the sky ~\citep{Stappers_2011}. For the HBAs there are two beamforming steps. An initial analogue beamforming takes place at the tile level, combining each group of 16 elements. Next the analogue Radio Frequency (RF) signals are transferred via coaxial cables to a Receiver Unit (RCU), where the data streams are amplified and digitally sampled by a 12-bit A/D (analogue to digital) converter. The clock operates at $200$~MHz; there is also a $160$~MHz option but this is infrequently used, does not apply to any of the data considered in this paper and is being deprecated in the LOFAR2.0 upgrade. In this work we Nyquist sample the $100-200$~MHz band. The digitized signal is fed to the Remote Station Processing (RSP) boards. This $100$-MHz wideband input signal is divided into $512$ sub-bands by a polyphase filter resulting in complex data with a $195.3125$~kHz subband width. A FIFO buffer compensates for differences in signal delays in the coaxial cables; this is termed the coarse delay. The next step of the RSP is combining and phase-rotating the signal, correcting for fine delays for each individual tile to form the station beam. Beamforming is performed independently for each sub-band and the products are called beamlets. It is possible to point each beamlet in a different location on the sky. The sampled values can be stored at either 16- or 8-bit depth, resulting in 244 or 488 beamlets due to hardware constraints. For this work, we point all beamlets at the same location so as to have effectively one beam with the widest possible bandwidth. Given that the loss in RFI discrimination performance is negligible and the 8-bit recording mode allows for a wider bandwidth, we record using 8-bit mode. One can then choose any $488$ beamlets from the $512$ that span the $100-200$~MHz band. In practice the lowest frequencies, in the FM band (which covers $87.5$-$108$~MHz range, \citet{Offringa_2013}), are omitted (see Sect.~\ref{sec:data} for data cleaning approach). An example of the signal power for the X and Y polarisation channels, averaged over all antenna, including all 512 sub-bands and integrated over $1$ second, is shown in Fig.~\ref{fig:bandpass}. The narrowband spikes represents the RFI contamination, clearly affecting the data quality at the beginning of the band.

\subsection{Single station sensitivity}\label{station_sens}

The sensitivity of a telescope such as LOFAR can be described using the radiometer equation~\citep{dicke46,vnr82,Handbook_2004}.
%\citep[for a detailed discussion see][]{Handbook_2004}.  
The mean flux density $S_{\rm mean}$, i.e. the integrated intensity of a pulse profile averaged over the whole pulse period, can be used to express the S/N of the folded pulse as \citep{Handbook_2004}:
\begin{equation}\label{eq:RIMEinv}
    SNR = \frac{G\,S_{\rm mean}\sqrt{n_{\rm p}\tau \Delta\nu}}{T_{\rm sys}}\sqrt{\frac{P-W}{W}}\;,
\end{equation}
where $G$ is the telescope `gain' or directivity ($G=4\pi^2A_{\rm eff}/\lambda^2$ for a single bow-tie antenna, with $A_{\rm eff}$ the effective collecting area), $n_{\rm p}$ is the number of polarizations, $\tau$ is the integration time, $\Delta\nu$ the frequency bandwidth, $W$ the pulsar profile width, $P$ the pulsar period and $T_{\rm sys}$ is the system noise temperature.

The $A_{\rm eff}$, in $\rm m^2$, of a single HBA dipole is modelled~\citep{Van_Haarlem_2013} as: 
\begin{equation}
    A_{\rm eff}^{\rm dipole}(\lambda)=\min \left\{\frac{\lambda^2}{3}, \frac{25}{16}\right\} =\min \left\{\frac{\lambda^2}{3}, 1.5625\right\}\;,
\end{equation}
with $\lambda$ being the observing wavelength, expressed in meters. The effective area cannot exceed the physical area, hence the $(5\times5)/16=1.5625\;\mathrm{m}^2$ limit. In this case, the gain is $4.2$, or $6.2$~dBi. The effective area for an array of $N=96\times16$ antennas is then modelled~\citep{Van_Haarlem_2013} simply as:
\begin{equation}
    A_{\rm eff}^{\rm max}(\lambda)=N* A_{\rm eff}^{\rm dipole}(\lambda)\;
\end{equation} 
although in reality there are known couplings between antennas which modify the situation and are not well quantified. $A_{\rm eff}$ can be represented via the Mueller matrix formalism\footnote{The Mueller matrix is a 4x4 matrix, relating the measured Stokes vector to the intrinsic one of the incident radiation. Derived from a combination of the Jones 2x2 matrix elements~\citep{SmirnovII_2011}. }, as the product of maximum effective collecting area $A_{\rm eff}^{\rm max}$ (frequency dependent) and the Mueller elements $m_{II}$ (see Sect.~\ref{sec_dreambeam}):\begin{equation}
    A_{\rm eff}=m_{II}A_{\rm eff}^{\rm max}\;.
\end{equation}

%However, the complex modelling of the LOFAR antenna beam requires an accurate evaluation of the system temperature.
The system noise temperature characterizes the combined noise contributions from the telescope itself and the sky \citep{bookEssentialRadio}. It comprises many factors, such as the receiver noise temperature, the spillover radiation from the ground and the contaminating sky emission, whereas the noise associated with the atmosphere is negligible at LOFAR frequencies. It can be summarized as: 
\begin{equation}
    T_{\rm sys}=T_{\rm ant}+T_{\rm sky}\;,
\end{equation}
with $T_{\rm ant}$ being the antenna temperature, which represents the instrumental noise in the absence of any incident sky signal, and $T_{\rm sky}$ is the sky brightness temperature, dominated by radio emission from the Galaxy and modelled as $T_{\rm sky}=T_{S0}\lambda^{-\alpha}$~\citep{Van_Haarlem_2013}, where $\alpha = 2.55$ ~\citep{Lawson_1997}. The sky temperature also includes a $2.725$~K contribution for the CMB emission, but negligible too.

\textsc{PyGDSM} interface can be employed to estimate the sky background noise ~\citep{Price_2021}. It incorporates different electron density models, as Haslam ~\citep{Haslam_1982}, GSM2008 ~\citep{deOliveira-Costa_2008}, GSM2016 ~\citep{Zheng_2017} and LFSS ~\citep{Dowell_2017}, which all consider the diffuse Galactic emission derived from different sky surveys at radio frequencies, along with different map fitting methods and model components.
The spectral index of the fitted power law, $\alpha$, varies with frequency (since the beam shape becomes wider at lower frequencies, different regions of the diffuse Galactic emission will affect differently the beam-averaged spectral index), sky region and sidereal time. For this work, antenna gain response is accounted for, and the antenna beam is computed via \texttt{DreamBeam} software as Jones correction factors (see Sect.~\ref{sec_dreambeam} and Sect.~\ref{sec:discussion}). 
%The antenna temperature is estimated as in \citet{Kondratiev_2016}, with a sixth-order polynomial fit. 

Accordingly, the sensitivity of a single LOFAR station depends on the source coordinates in the sky and on the observing frequency; the latter effects the effective area but also the beam shape. These factors combine to impact the instrumental sensitivity, leading to significant degradation below $\sim 30^{\circ}$. 
The dependence of $A_{\rm eff}$ alone on elevation-driven sensitivity variations due to projection effects can be approximated by a cosine-squared relationship, with the effective collecting area decreasing as $\sim \cos^2(ZA)$, where $ZA$ is the zenith separation (see Malus's law e.g. ~\citet{Optics_2017}). However the full response of individual LOFAR stations is more complex and it has been studied in several previous works \citep{Noutsos_2015,Kondratiev_2016,Blaszkiewicz_2018}. \citet{Noutsos_2015} used pulse profiles from four bright pulsars, to demonstrate that the sensitivity of LOFAR's 24-core stations, as a function of elevation, decreases as $\cos^{-1.39}(ZA)$. This study showed that the sensitivity of the LOFAR core array is more than $50 \%$ greater than the case of uncalibrated data (i.e. without correction for direction-dependent effects), where the rms noise would decline as $\sin^{-2}(el)$.  A comparable study involving only one station was conducted using the PL612 HBA antennas in Poland: \citet{Blaszkiewicz_2018} fit the rms noise of their observations and found that it is best fit by the function $0.1/\rm cos^5(ZA)$, while noting a practical minimum elevation of 25 degrees. 

Effects at antenna level will affect the station beam, which adds to the complex modelling of the spectral, temporal and spatial variation of the beam itself. Sources outside the main lobe can cause a defect in the calibration solutions and the presence of damaged antennas modify the overall tile (and station) beam. At any given time a non-zero number of the $96\times16\times2=3072$ antennas are typically malfunctioning/not operational, meaning various non-uniform weightings must be accounted for. For example, \citet{Brackenhoff_2025} developed a python package to simulate the effects of malfunctioning antennas in HBA tiles: when at least $9\%$ of a HBA tile is down, the beam shape is perturbed and the errors propagate to the combined station beam. The effect is mainly seen in the region of `beam-nulls' and sidelobes, where the different beam pattern could lead to a non-negligible contamination from A-Team sources (i.e. highest flux density in the northern sky) to the primary beam.

\subsection{\texttt{DreamBeam}}
\label{sec_dreambeam}
Modelling the station beam constitutes a major step in the signal processing pipeline. Formally, in the case of a phased-array telescope, a virtual antenna is generated by the implementation of a phase delay for the signal coming to the different elements. In doing so, a station beam is formed. The station beam is encoded as a factor in the calibration procedure, usually computed via electromagnetic simulations and tested through dedicated observational measurements ~\citep[e.g.][]{Mol_2011, Yatawatta_2012, Van_Haarlem_2013}. Calibration procedure takes care of several systematic effects acting on the electromagnetic waves from the source to the output of the interferometer. Signal correction is made through the implementation of the RIME, as an extension of the fundamental van Cittert-Zernike theorem~\citep{Carozzi_2009}. The RIME formalism has been revised multiple times \citep{Sault_1996, Nijboer_2007, SmirnovI_2011, SmirnovII_2011, SmirnovIII_2011, SmirnovIV_2011, Tasse_2012, Price_2015, Murley_2019, Creaner_2019, Kansabanik_2025}. 

When the electromagnetic signal arrives at the antenna, it is converted into complex voltages (as described above in Sect.~\ref{station_bf}). The relation between the incident signal and the voltages as output of the interferometer can be represented as a linear operation between vectors, driven by a matrix $J$, the Jones matrix (derived from the Jones calculus adopted for optical systems ~\citet{Jones_1941}). In principle, the $J$-matrix accounts for both direction-dependent and direction-independent effects, such as antenna phase differences, Faraday rotation, parallactic angle changes, feed responses and the complex gain factor of the electronics \citep{Hamaker_1996}. The $J$ matrix is expanded in a `Jones chain' as defined in \citet{SmirnovI_2011}. The chain translates in a multiplication of several matrices, each one represents an instrumental/environmental effect. The order in the `chain' follows the sequence of the effects acting on the signal as it propagates from the source to the observer. Using this mathematical framework, the correction factors are antenna-, direction-, time- and frequency-dependent.
To evaluate the RIME and associated Jones matrices for a given LOFAR station, the \texttt{DreamBeam} software was developed. \texttt{DreamBeam} is a versatile python-based pipeline, able to model the telescope response and predict the beam pattern of a LOFAR station~\citep{Carozzi_2016}. \texttt{DreamBeam} provides the output instrumental response, namely the Jones matrix elements: $J_{\rm XX}$, $J_{\rm XY}$, $J_{\rm YX}$ and $J_{\rm YY}$ for input sky coordinates and observing frequency. The correction factors are: the so-called \textit{P} matrix accounting for the projection from a celestial centred frame to the topocentric-station-frame and the \textit{E} matrix accounting for the antenna polarization response. The beam model is based on the formalism developed by Hamaker et al. \citep{Hamaker_1996, Hamaker_2000, Hamaker_2006}. Using \texttt{DreamBeam} the response of the single polarisation channel can be evaluated as $J_{\rm XX}^2+J_{\rm XY}^2$  for the X-channel and $J_{\rm YY}^2+J_{\rm YX}^2$ for the Y-channel (see Sect.~\ref{sec:discussion} for further details).  

From the Jones elements, the Mueller matrix factors, introduced in Sect.~\ref{station_sens}, can be obtained as~\citep{Kondratiev_2016}: 
\begin{equation}
    \rm m_{\rm II}(\rm f, \rm ZA, \rm az)=\frac{1}{2} (J_{\rm XX}J_{\rm XX}^*+J_{\rm XY}J_{\rm XY}^*+J_{\rm YX}J_{\rm YX}^*+J_{\rm YY}J_{\rm YY}^*),
\end{equation}
with $\rm m_{\rm II}$ factors as function of frequency ($\rm f$ ), zenith angle ($\rm ZA$) and azimuth ($\rm az$).
Mueller matrices predict the time and frequency evolution of the effective collecting area for a given pointing direction. Based on the adopted beam model, it relates the measured and the `true'/intrinsic Stokes vector of an observed source. Hence, it represents a correction factor tracing the changes in the station sensitivity. 

The performance of \texttt{DreamBeam} was tested by \citet{Creaner_2019}. It was noted that apparent polarization and flux density variations can result from misinterpretations of the intrinsic response of the antennas as a source is tracked across the sky, varying in both altitude and azimuth. By comparing observed data of Cassiopeia A (CasA) with simulated observations generated at the same station, they reported non-negligible flux density variations (with the discrepancy being more pronounced in X-polarisation channel compared to the Y-polarisation channel). These variations were found to be dependent on 4 main parameters: frequency, time, altitude/elevation and azimuth \citep{Creaner_2019}. 
\begin{figure}[H]
   \centering
   \includegraphics[width=0.9\hsize]{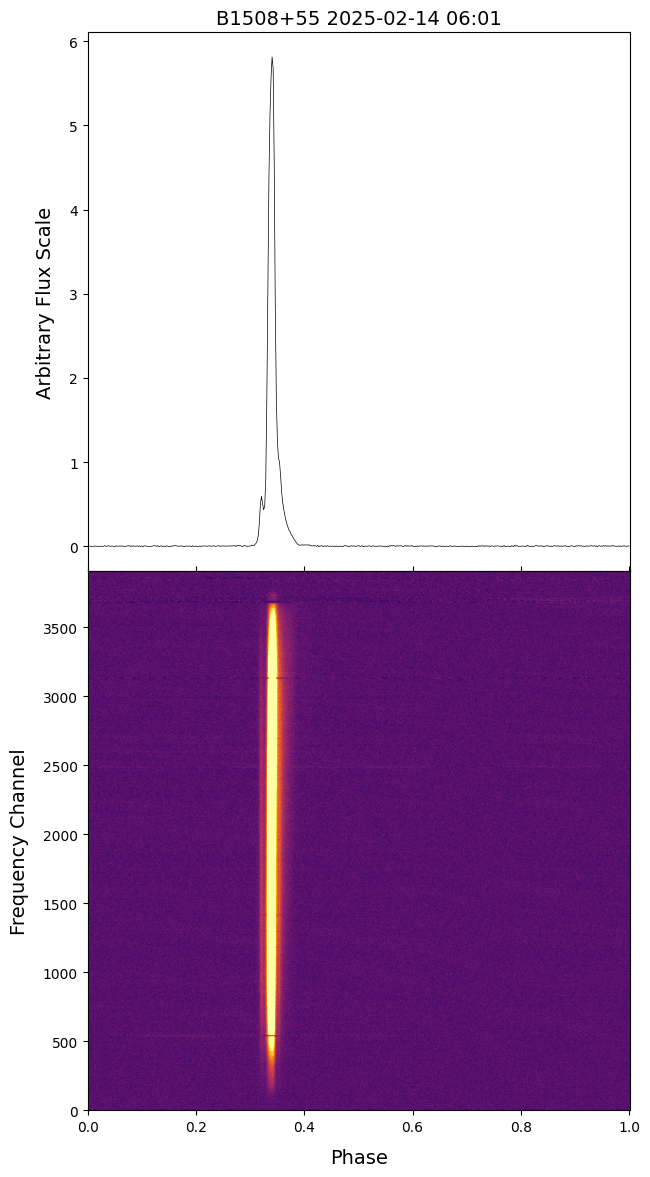}
\caption{The folded pulse profile as intensity over pulse phase is shown at the top, while the phase-frequency plot is shown at the bottom, with frequency channels displayed on the y-axis. The profile shown here is RFI-cleaned, hence some frequency channels appear removed. As an example, we used the integrated pulse profile of PSR B1508$+$55 for the $59$ min-observation taken at $06:01$ UTC on $14$ February $2025$, which corresponds to $87.6^\circ$ in elevation. The retrieved S/N is $4048.59$, or more formally stated: $4049(1)$.}
\label{fig:pulseprofile}
\end{figure}

\vspace{5mm}

\begin{table*}[!t]
\caption{Summary of the main pulsar properties and track monitoring dates and times.}        
\label{table:pulsarsI-LOFAR}      
\centering     
\resizebox{\textwidth}{!}{
\begin{tabular}{c c c c c c c}        
\hline\hline                
PSR & RA(hh:mm:ss.s) & DEC(dd:mm:ss) & Start Time & End Time & DM ($\mathrm{pc\,cm^{-3}}$) & DM uncertainty ($\, 10^{-4}$)\\   
\hline   
   J2145-0750 & 21:45:50.45 & -07:50:18.59 & 2024-12-05 14:00 & 2024-12-05 20:00 & 9.00441 & 0.6\\
   B0149-16 & 01:52:10.85 & -16:37:53.64 & 2025-01-31 14:00 & 2025-01-31 18:00 & 11.92707 & 3.3 \\
   B0950+08 & 09:53:09.31 & +07:55:35.75 & 2024-10-03 03:01 & 2024-10-03 15:01 &2.96898 & 1.0\\ 
   B1133+16 &11:36:03.12 & +15:51:14.18 & 2024-11-19 01:30 & 2024-11-19 14:30 & 4.83998 &1.2\\
   B2016+28 & 20:18:03.83 & +28:39:54.21 & 2024-11-04 13:31 & 2024-11-04 22:40 &14.20164 & 0.5\\
   B1946+35 & 19:48:25.01 & +35:40:11.06 & 2024-11-22 06:30 & 2024-11-23 00:30 & 128.86797 & 197.1\\
   B2217+47 & 22:19:48.14 & +47:54:53.93 & 2024-10-03 17:01 & 2024-10-04 08:01 & 43.49760 & 0.8\\ 
   B0329+54 & 03:32:59.41 & +54:34:43.33 & 2024-01-09 08:00 & 2024-01-10 14:20& 26.76745 & 2.4\\
   B1508+55 & 15:09:25.63 & +55:31:32.39 & 2025-02-13 11:01 &2025-02-14 10:01 & 19.61884 & 0.6\\
   B2224+65 & 22:25:52.86 & +65:35:36.37 & 2025-01-21 16:30 & 2025-01-22 15:30 &36.44574 & 5.3\\
   B0809+74 & 08:14:59.50 & +74:29:05.70 & 2025-02-04 13:30 & 2025-02-05 11:30 & 5.75073 & 2.8\\
\hline                                 
\end{tabular}}
\end{table*}
\newpage
\section{Data acquisition and data processing}\label{sec:data}

A total of $11$ pulsars were observed for this study (Table \ref{table:pulsarsI-LOFAR}). The targets are distributed in declination from $\sim$ -10$^{\circ}$ to +80$^{\circ}$. This range enables a broad characterization of the response of the instrument, both near the zenith and closer to the horizon, where several effects begin to degrade the S/N.\\
PSRs~B2217$+$47, B0329$+$54, B1508$+$55, B2224$+$65 and B0809$+$74 are circumpolar sources, i.e. their declination is larger than $(90-\phi)$ where $\phi$ is the station's latitude ($\phi = 53.09^\circ$ for IE613). As such they never go below the horizon at Birr. Using \textit{these} pulsars, we are able to cover full uninterrupted tracks. For lower declination pulsars we simply cover as much of the above-horizon trajectory of each source as possible, also given other operational constraints at the Observatory. The source selection was made based on each pulsar's declination and flux density, all the PSRs in the sample are well known bright pulsars (see \href{https://www.atnf.csiro.au/research/pulsar/psrcat}{ATNF pulsar catalogue},~\citet{manchester_ATNF_2005}). Each target was observed for several hours, in contiguous 1-hour blocks (for operational reasons) consisting of $59$~min on source with $1$-min gaps. The observations went across a sidereal day or two (see column \textbf{$(4)$} of Table~\ref{table:pulsarsI-LOFAR}).
The data were recorded with the HBAs in local mode in the frequency range 110-190 MHz (from 12 to 499 subbands), which corresponds to RCU mode $5$. The observational parameters are reported in Table~\ref{tab:obs_param}.\\
The data are processed on the REAL time Transient Acquisition compute cluster \citep[][REALTA]{Murphy_2021}. Once the raw voltages are collected, they are reduced to 32-bit Sigproc Stokes $I$ filterbanks~\citep{Lorimer_2011}, using \textsc{udpPacketmanager}~\citep{udpdavid2023}. Coherent dedispersion is then applied in order to correct for dispersion due to the interaction between the incoming signal and free electrons in the ISM along the line of sight, by means of the \textsc{cdmt} software~\citep{Bassa_2017}. 
Subsequently, the filterbanks are carefully downsampled to 8-bit Sigproc filterbanks using the \textsc{digifil} tool~\citep{Van_Straten_2011}. Science-ready data products are then produced in the form of archive files generated by \textsc{DSPSR}~\citep{Van_Straten_2011}, an open source C++ library for digital signal processing. The final processed data have a time sampling of $81.92\, \mu \rm s$ and frequency resolution of $24.4$ kHz. Each 1-hour pointing is folded separately using the timing solution obtained from the ATNF pulsar catalogue~\citep{manchester_ATNF_2005}. In the post-processing phase, each observation was cleaned of radio frequency interference using a modified version of the \textsc{CoastGuard} software package (see \citet{Lazarus_2016}, \href{https://www2.physik.uni-bielefeld.de/fileadmin/user_upload/radio_astronomy/Publications/Masterarbeit_LarsKuenkel.pdf}{Kuenkel, in prep.}\footnote{\url{https://github.com/larskuenkel/iterative_cleaner}}). Starting from $3904$ channels in frequency, a given number (around a few percent for all pulsars) were masked per observation. We used the methods outlined in \citet{Susarla_2025} to obtain the dispersion measures (DMs) and subsequently dedisperse each folded archive with the obtained DM. The values so-determined are listed in Table ~\ref{table:pulsarsI-LOFAR}. After updating the DM, we used the \textsc{pdmp}~\citep{Hotan_2004} tool to obtain the S/N, upon time and frequency scrunching. For each pulsar the values of the measured S/N at each sampled altitude and azimuth are reported in Appendix ~\ref{app:A}. A typical example of a cleaned folded 1-h pulse profile for PSR~B1508$+$55 is shown in Fig. ~\ref{fig:pulseprofile}.
\begin{table}[h!]
    \caption{Parameters of observational data.}
    \centering
    \begin{tabular}{l l l}
       \hline\hline
     
       \hline
        Parameter & IE613 stn & SE607 stn\\
        \hline
        Duration &  $3600$ s & $3600$ s\\
        Lower Frequency & 102.3 MHz & 109.6 MHz \\
        Central Frequency & 149.9 MHz & 149.7 MHz\\
        Upper Frequency & 197.5 MHz & 189.9 MHz \\
        \#Frequency channels & 3904 &  3296\\
        Bandwidth & $95.2$ MHz & $80.3$ MHz\\
        Channel width & 24.4 kHz & 24.4 kHz\\
        \hline
    \end{tabular}
    \label{tab:obs_param}
\end{table} 
\section{Results}\label{sec:results}

\begin{figure*}[t]
  \centering
    \includegraphics[scale=0.6]{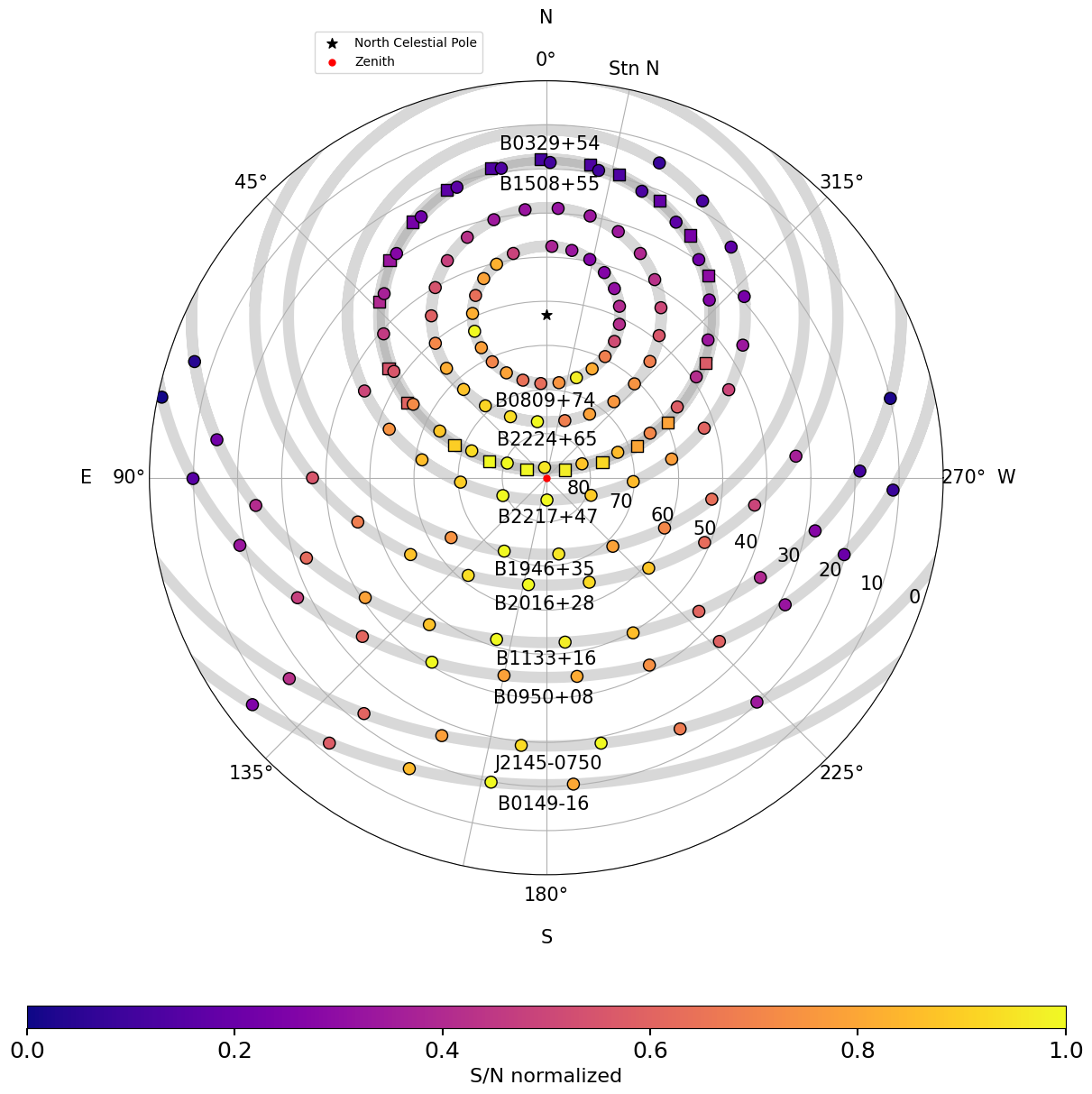}
  \caption{Synoptic plot of pulsar observations tracking their path as they cross the I-LOFAR sky. The colour-coded S/N is normalized to the maximum S/N for each pulsar. Grey shadowed lines highlight the target path. PSR B0329$+$54 is shown with squared markers in order to differentiate it from PSR B1508$+$55, because of the overlapping of the relative tracks, at the resolution shown. Note: each pointing lasts 1 hour; here the points are located at the \textit{first} instant of each observation. Asymmetry in the S/N between the rising and setting sides of the pulsar trajectories is visible by eye in this plot. `Stn N' labels the North direction in the reference frame of the dipoles.}
  \label{Fig.synoptic_plot}
\end{figure*}

Fig.~\ref{Fig.synoptic_plot} shows the retrieved S/Ns for all 11 pulsars as a function of sky position. Each grey curve represents a target track, with pointings colour-coded by S/N, normalized to the maximum for each target. Through high-precision DM measurements, the retrieved S/N values range from several thousands down to below ten for the very lowest altitudes (see Table~\ref{table:appA} in Appendix \ref{app:A}). Quoted S/N values are obtained with the \texttt{pdmp} tool from the \textsc{psrchive} software suite~\citep{Hotan_2004}. This optimises the period and DM parameters to get the highest S/N value; we did not employ DM search techniques of the kind commonly used in Fast Radio Burst (FRB) studies which optimise for structure~\citep{Amiri_2024, Chawla_2025}. Some interesting effects dominate the integrated pulse profile for PSR~B0809$+$74, where interstellar scintillation can be seen from its dynamic spectra \citep[for further details, see previous observations as in][]{Basu_2023, Wu_2022, Hassal_2012, Ricket_2000}. 
\\
Hence, extra caution is advised when interpreting the measured values for this pulsar (see last panel of Fig.~\ref{fig:all_snr_azimuth} of Appendix~\ref{app:B} and Table~\ref{table:appA}) as there is clearly non-negligible effects that are neither due to intrinsic pulsar properties of those of our instrumentation. For each pulsar the best response is consistently reached near the highest altitude, with overall higher values closer to the zenith (i.e. the centre of the synoptic plot).

Plots of S/N versus elevation, colour-coded by azimuth, are displayed in Fig.~\ref{fig:all_snr_azimuth} in Appendix~\ref{app:B}. The sensitivity significantly drops at lower elevations, with a decrease of at least one order of magnitude compared to higher elevations. We note that the decrease in S/N is substantial for all the targets, corresponding to a reduction ranging from $\sim 70\%$ and $\sim 90\,\%$ between zenith observation and lower elevations. When the tracks are circumpolar (e.g. PSR B1508+55 or PSR B2224+65) a pattern reminiscent of a hysteresis curve is clearly visible. Especially for these objects, an overall better response is achieved when the target is rising (lower azimuth) towards its upper culmination as opposed to when its elevation is descending (i.e. higher azimuth).
\begin{figure}[h]
   \centering
   \includegraphics[width=\hsize]{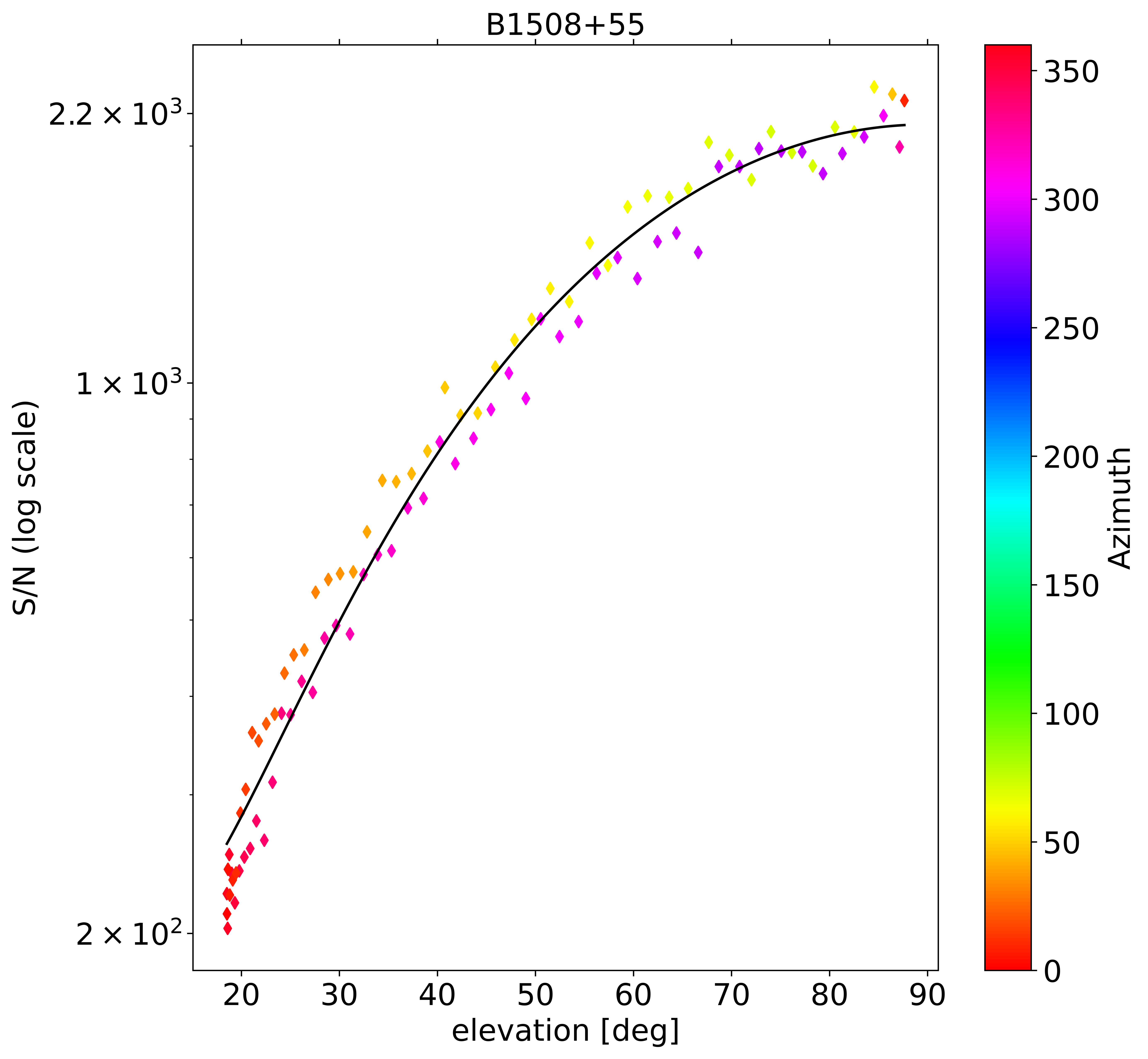}
      \caption{S/N versus elevation in semi-log scale for PSR B1508+55 and for the sliced observations of $15$ minutes. The colour bar refers to the azimuth coordinate. The solid black line represents the fitting with $\cos^2(ZA)$.}
         \label{Fig.B1508+55_S/Nvsel_15min}
\end{figure}
\begin{figure}[h]
   \centering
   \includegraphics[width=\hsize]{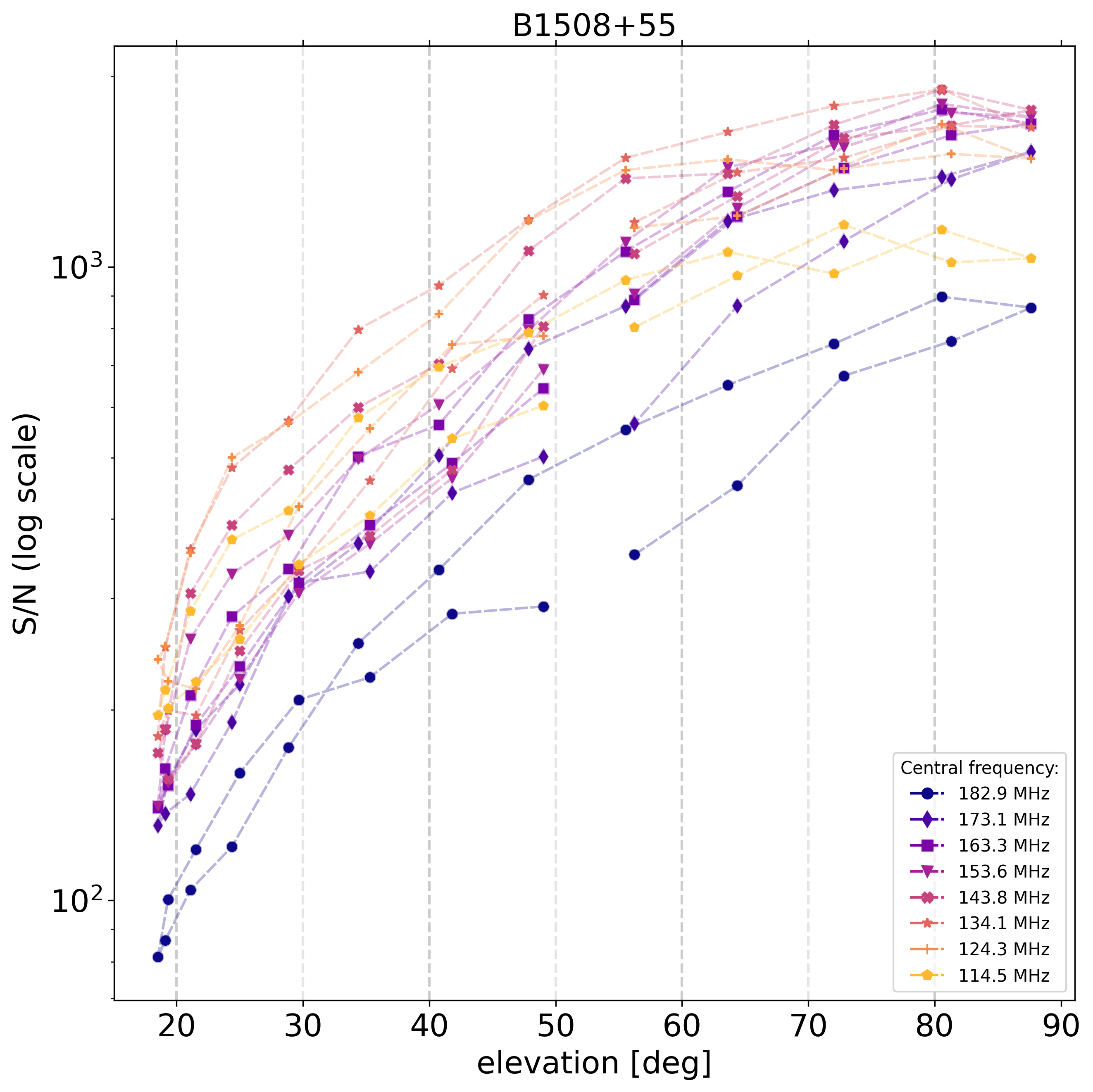}
      \caption{S/N versus elevation in semi-log scale for PSR B1508+55. Each curve represents a slice of $\sim 10$ MHz in bandwidth.}
         \label{Fig.B1508+55_S/Nvsel_10MHz}
\end{figure}\\
We next endeavoured to measure the variation of the station performance in time and frequency at finer resolution. The shorter integration time might reveal variations in S/N, hidden otherwise in the 1-hour averages. In this way, the elevation dependence of the antenna response can be examined in greater detail. We first sliced the observations in time, using PSR~B1508$+$55 as a test. We cut the observations in time by a factor of $4$, resulting in $15$-minute pointings. This reduces the S/N by a factor of $\sim2$ but the signal is sufficiently strong that this is not a serious issue in the case of PSR~B1508$+$55 (and a few others). The shorter time scales reveal the same hysteresis-like pattern as previously highlighted, but with the increased granularity we can see  the maximum S/N is reached closer to $90^\circ$ compared to the longer pointings. We also see higher values of S/N for the rising part of the curve (corresponding to smaller azimuths). The result is shown in Fig.~\ref{Fig.B1508+55_S/Nvsel_15min}, where the S/N is plotted in semi-logarithmic scale with the elevation on the x-axis. In this plot, we also show the fit of the data points with a $\cos^2$ function of the ZA; the station response is better (worse) than this model while the source is rising (setting).
Using PSR B1508+55 again, we cut the observations in frequency. Each pointing is now sliced in $\sim 10$-MHz sub-bands, comprised of $400$ channels. We use the same procedure for the post-processing, and we retrieve the S/N, which is shown in Fig.~\ref{Fig.B1508+55_S/Nvsel_10MHz}.
The measured S/N display the same general trend observed for all pulsars with IE613. The measured S/N, in each sub-band, preserve the same asymmetry in azimuth, with the rising part of the track showing an overall better sensitivity of the station. Additionally, for this pulsar the frequency dependence shows that the best response is achieved for the $\sim 129 - 139$~MHz sub-band with central frequency of $134.1$~MHz. This is due to a combined effect of the effective collecting area and the sky background noise: S/N scales as $A_{\rm eff}/T_{\rm sys}$, where the $A_{\rm eff}$ falls off with $\nu^2$, with $\nu$ the observing frequency, and $T_{\rm sys}$ increases as the observing frequency decreases. Adding the fact that the observed S/N is given by the convolution of the ratio $A_{\rm eff}/T_{\rm sys}$ and the intrinsic pulsar spectrum, this leads to a peak in response around $135$~MHz \citep{Wijnholds_2011}.
\begin{figure}[h]
   \centering
   \includegraphics[width=\hsize]{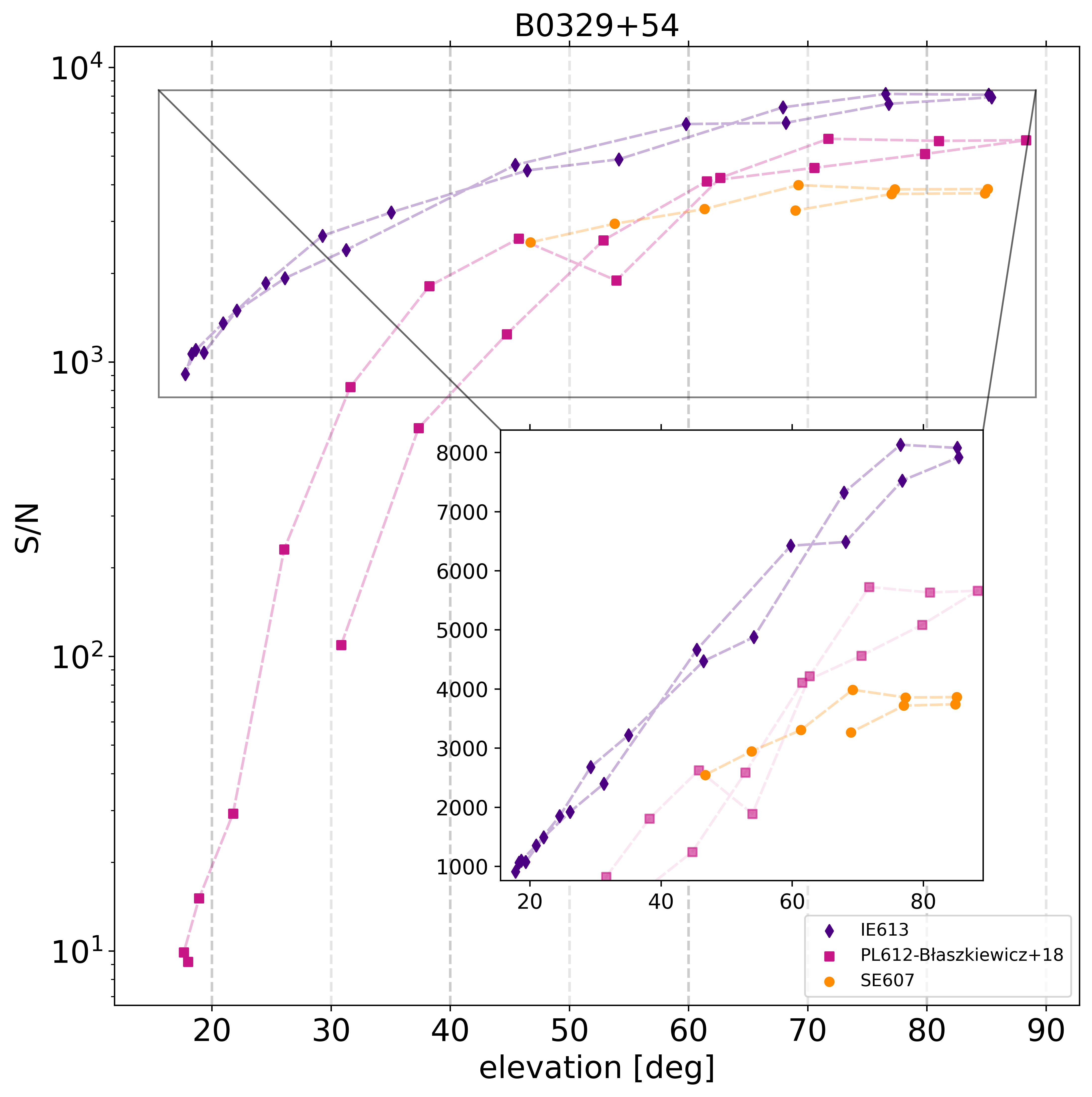}
      \caption{S/N of PSR~B0329+54 as observed with 3 set of HBAs, with the stations located in Ireland (dedicated observations), Poland \citep{Blaszkiewicz_2018} and Sweden (dedicated observations) respectively. The zoomed panel shows the S/N above $1000$ in linear scale. At similar elevation, the rising phase of PSR~B0329+54 is characterised by a higher S/N.}
         \label{Fig.IE613vsPL612vsSE607}
\end{figure}
For PSRs~B0329+54 and B1508+55, we compare the observed S/N from this study with the S/N obtained by \citet{Blaszkiewicz_2018}. At PL612, PSR~B0329+54 was targeted for a total of $27$ hours on the 19--20 May 2017, while PSR~B1508+55 was observed for 10 hours on the 12-13th of May. We further observed PSR~B0329+54 with the SE607 station, using the HBAs. The pulsar was targeted for $9$ hours, across the upper culmination point. The data were processed as described in Sect.~\ref{sec:data}. In Fig.~\ref{Fig.IE613vsPL612vsSE607}, we show the observed S/Ns for the $3$ stations. PSR B0329+54 is an ideal target because it is bright and circumpolar for all $3$ observatories, allowing its tracks to cover the full range of azimuth and altitude. The plot suggests that the trend is preserved for the same target and same frequency range of the observations. Overall, in each case the rising track displays a higher S/N compared to the setting curve. The overall better performance achieved at IE613 in comparison to SE607 can be explained by the wider bandwidth used at IE613. For PL612 there is the additional deleterious effect that the data were not coherently dedispersed, which leads a precarious drop in response by relative factor $\sim100\times$ as the source approaches the horizon.

A summary of our observations is that the response of IE613 (and SE607 and PL612) is asymmetric in azimuth such that every pulsar we observe is seen more significantly as it rises, rather than when it is setting. This is true across the frequency band. The magnitude of the effect is large, with responses often being $\sim20\%$ better on the rising side. The Hamaker beam model, as implemented in \texttt{DreamBeam} for example, predicts a much more symmetric response (see Sect.~\ref{sec:discussion}). The asymmetric response is also seen to be in least in qualitative agreement with imaging observations of Cas A performed by \citet{Creaner_2019}. 

\section{Discussion}\label{sec:discussion}
\begin{figure}[h]
   \centering
   \includegraphics[width=\hsize]{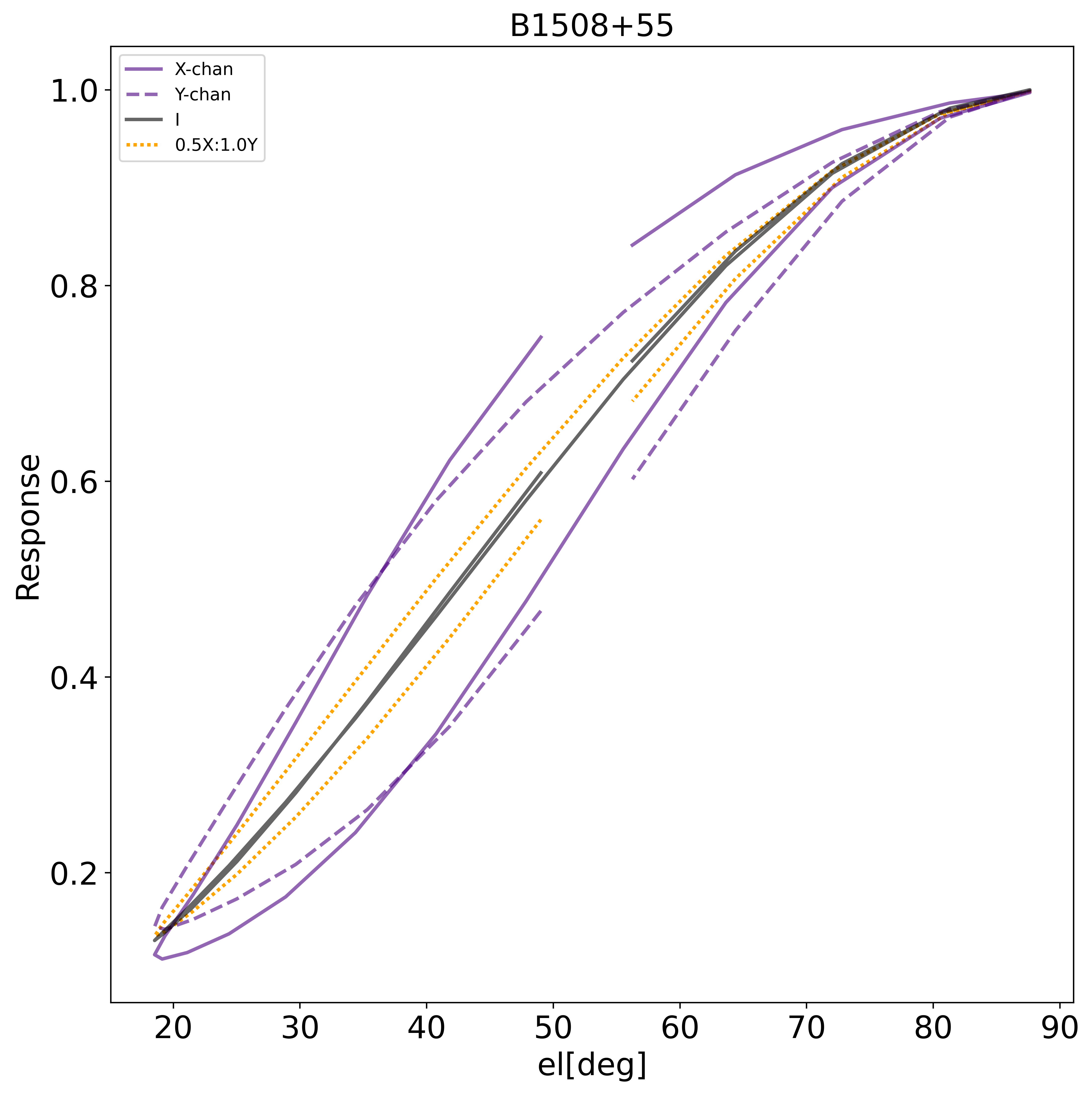}
    \caption{The expected response of the X and Y components of $m_{\rm II}$, obtained from the Jones coefficients generated through \texttt{DreamBeam}. The solid gray line represents the total power for equal weighting of X and Y, showing a highly symmetric response. Purple solid and dashed lines are the single channel components (meaning power in the X polarization channel and Y-polarization channel). The dotted orange line shows the combined Stokes I response. A representative unequal weighting is shown in dotted orange.}
         \label{Fig.XY_B1508_obs}
\end{figure}
Using pulsar observations, we have been able to characterize the response of the HBAs at the IE613 station: we have determined how the station performance varies with both altitude and azimuth through extended monitoring of $11$ pulsars as they cross the local sky. We have obtained S/N values, i.e. effectively $A_{\rm eff}/T_{\rm sys}$ values, as a function of elevation and azimuth. The sensitivity is given by the combined effect of the gain of the telescope and the multi-component noise. 

Therefore, we investigate the expected station response as a function of the horizontal coordinates according to the beam model implemented in \texttt{DreamBeam}. With this we evaluate the elements of the Jones matrices (introduced in Sect.~\ref{sec_dreambeam}), used in the RIME formalism. Here we present the modeled responses of the X and Y polarisation components, and the combined response adding these components, for IE613. We note that this gives us a measure of $A_{\rm eff}$, which trace the element beam pattern, rather than $A_{\rm eff}/T_{\rm sys}$. However we note that, except near the horizon where spillover and beam distortion can be important, the $T_{\rm sys}$ factor is more or less constant for the majority of the trajectory of each source.

To generate the Jones matrices, the input parameters include the pointing direction (the pulsar RA and Dec; here we use the coordinates of PSR~B1508$+$55), the station ID (defining its location and orientation), the observing band (in this case the HBA band) and the frequency at which to do the calculation ($150$~MHz is shown here). We evaluate these elements at $1$-hour intervals over a period of $24$ hours. Fig.~\ref{Fig.XY_B1508_obs} shows the single-channel components (purple solid and dashed lines) and the total power output of the antennas: $m_{\rm II}=XX^* + YY^*$ where $X=J_{xx}J_{xx}^*+J_{xy}J_{xy}^*$ and $Y=J_{yx}J_{yx}^*+J_{yy}J_{yy}^*$. The $X$ and $Y$ responses are such that when combined with equal weighting, the antenna pattern is symmetric, as is the resulting beam response
in azimuth. While some small degree of asymmetry is expected due to some projection and beam-shape related effects, the magnitude of this asymmetry is expected to be small and cannot account for the hysteresis-like pattern as in Fig. ~\ref{fig:all_snr_azimuth} in Appendix ~\ref{app:B}.

Fig.~\ref{Fig.XY_pointing_sim} illustrates how the dipole response varies as a function of the source declination at a given observing site (IE613 in this case). Here we simulate the response of the antennas considering targets with the same right ascension but spread at $1$ deg in declination. The simulation traces the targets in their rising/setting tracks over IE613. As an example, we report the simulated antenna gains of $\rm DEC=10^{\circ}$ (left panel) and $\rm DEC=65^{\circ}$ (right panel). The plots show the X and Y channel responses as given by \texttt{DreamBeam}, so displaying the Jones factors. For all `high declination' targets ($\rm DEC \geq 50^\circ$), the Y (X) component results higher on the rising (setting) part of it, similar to the right panel. For targets at smaller declinations, the X (Y) component results higher on the rising (setting) part of the tracks. However, as the declination angle approaches and crosses $\rm DEC = +37^\circ$ (above which sources are circumpolar at Birr),
\begin{figure*}[t!]
   \centering
   \includegraphics[width=0.47\hsize]{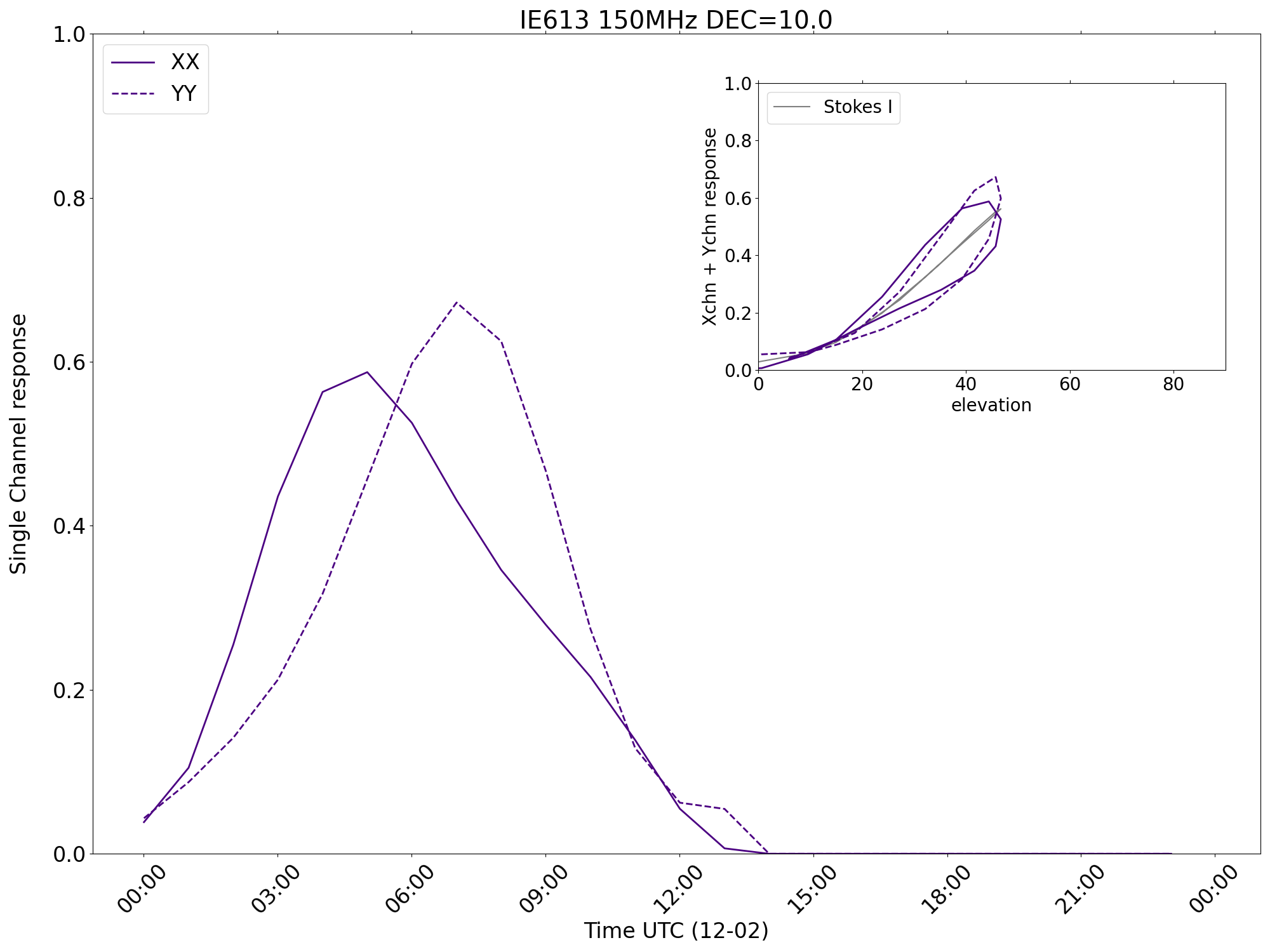}
   \includegraphics[width=0.47\hsize]{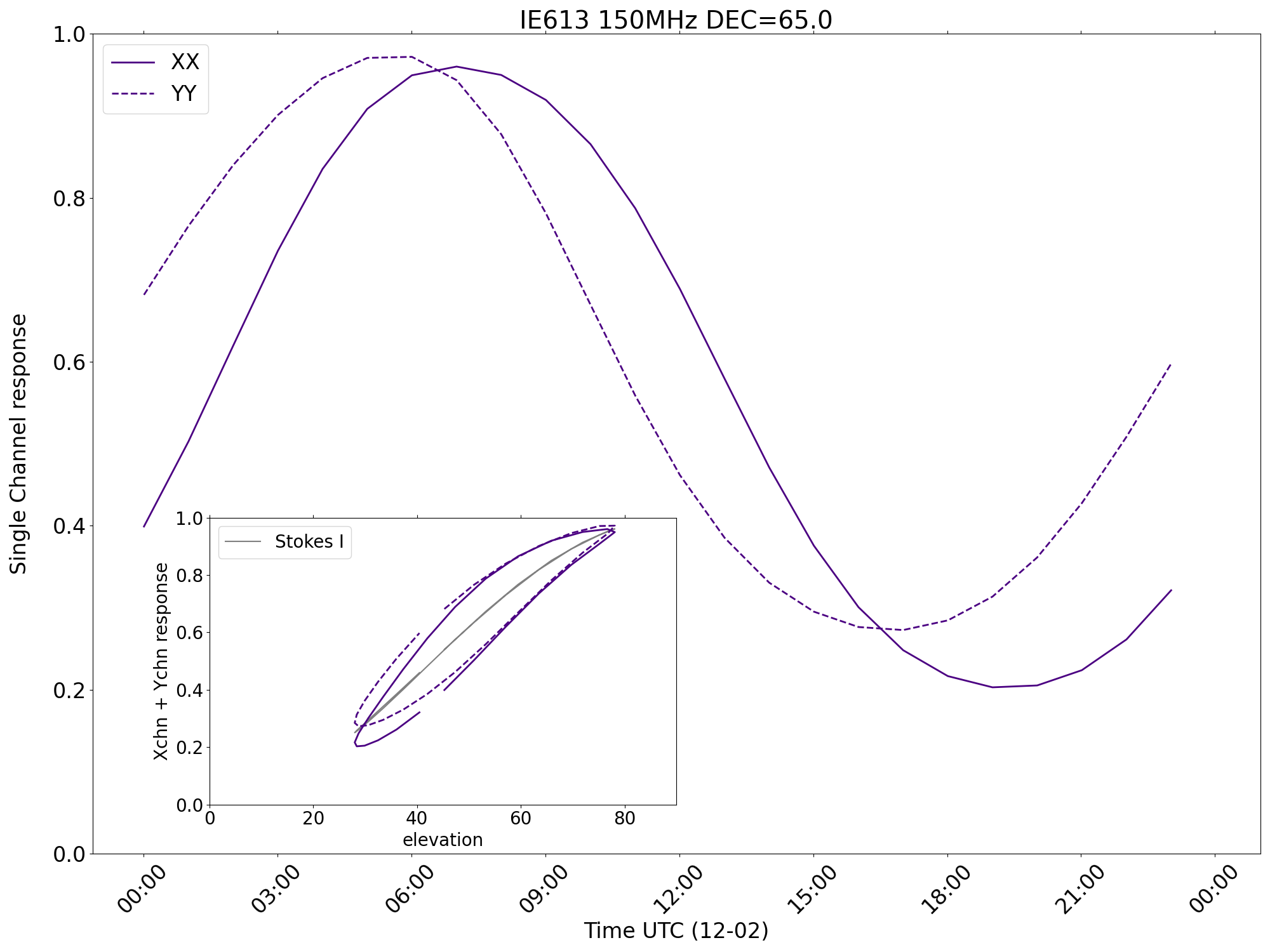}
      \caption{Modelled response through \texttt{DreamBeam} showing the variation in time. The response with respect to elevation is displayed in the inset panels. The simulation is based on observations at the IE613 station of a target with $\rm DEC=10^\circ$ (left panel) and $\rm DEC=65^\circ$ (right panel). The Jones factors shown here are generated at $150$~MHz. A Jones factor equal to $0$ indicates that the target is situated below the horizon.}
         \label{Fig.XY_pointing_sim}
\end{figure*}
the channel response becomes more complex and the two contributions are no longer clearly separated on the rising and setting parts of the track. An analogous trend has been observed when simulating the Jones elements and hence the beam patterns for the Swedish station, where the critical declination for circumpolar sources is defined at $\rm DEC = +33^\circ$. This behaviour reflects the relative antenna orientation across different stations. The dipoles are all aligned to the local north at the reference station, CS$002$ in Exloo (The Netherlands). It means that the dipoles are rotated by a given angle within the station to align to CS002. The East-West distribution of LOFAR stations across Europe isn't large enough to observe a $+90^\circ$ rotation of the dipoles with respect from one end to the other.\\
If such a rotation were achieved, the X and Y responses would at some point swap, with the X (Y) component being higher on the rising (setting) part of the source trajectory.

Given Eq. ~\ref{eq:RIMEinv}, the azimuthal asymmetry observed in our pulsar sample (see Fig.~\ref{fig:all_snr_azimuth}) can be attributed to two distinctive factors. One factor is the beam model, where this asymmetry can be explained by an imbalanced weighting of the X and Y polarisation components, thereby affecting $A_{\rm eff}$. If so, the origin of this effect remains unclear: as this behaviour is seen across IE613, SE607 and PL612 it may be systematic and so be the case in all international stations, so that individual local explanations do not suffice. 

\begin{figure}
    \centering
    \includegraphics[width=\hsize]{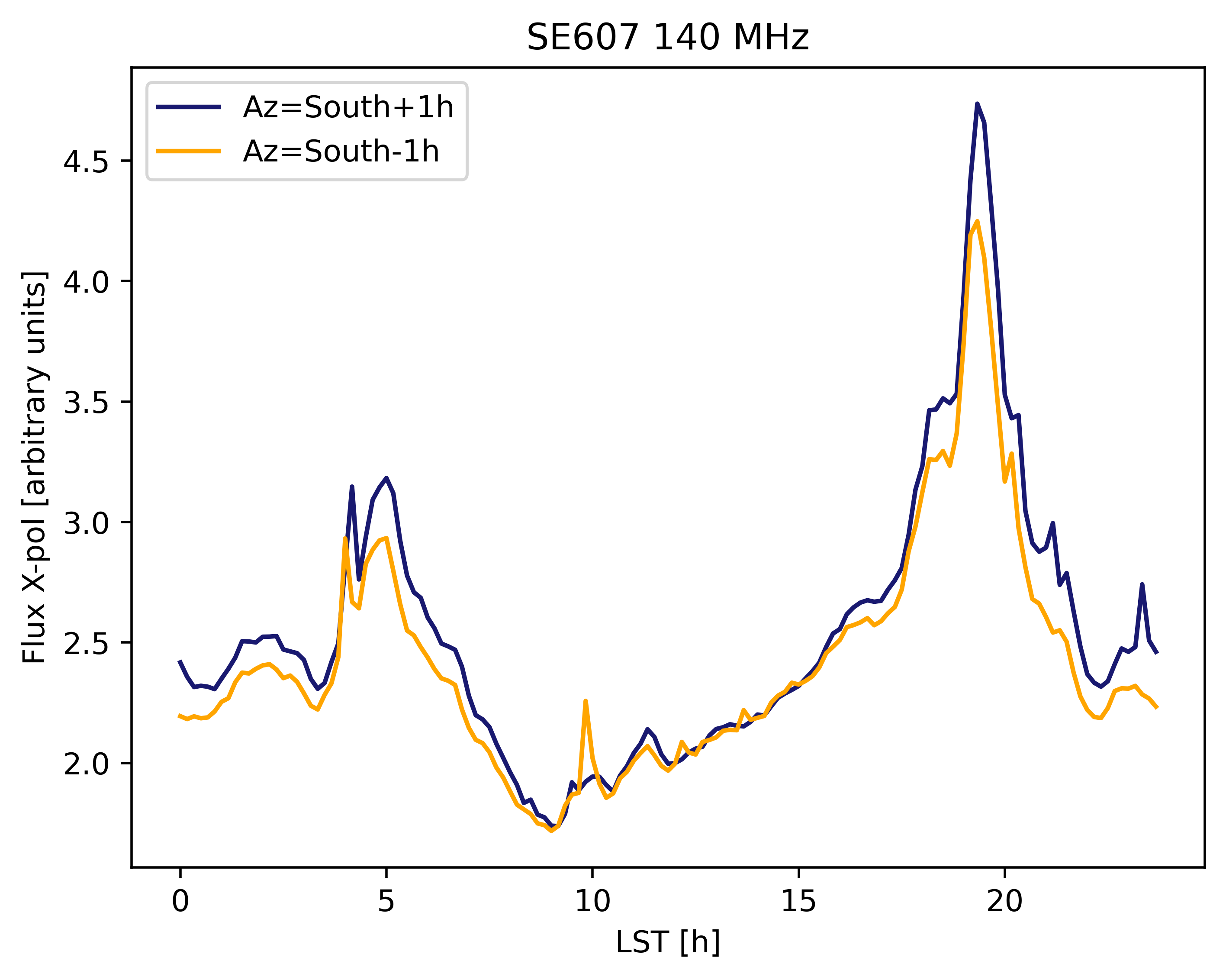}
    \caption{Test observations at $140$~MHz from the SE607 station to evaluate the sky noise variation. Two 24-hours pointings were performed, each at $\rm EL = 60^\circ$, and either $15^\circ$ ahead or behind of the local Southern meridian in azimuth. The blue (orange) curve is shifted +1hour (-1hour) in LST. In this way, the observations lines up along the time axis, showing how the response varies on the sky but on opposite sides of southern meridian.}
    \label{fig:tobia}
\end{figure}

The asymmetry is likely to be due to some standard configuration that is uniform across all stations. To choose one example: it could be related to the fact that the X-port for all LOFAR HBAs is also used for power supply to the low-noise amplifier \citep{Cookbook_2013}.
Fig.~\ref{Fig.XY_B1508_obs} provides an example of an asymmetric weighting, with a ratio of 0.5:1.0, i.e. $(aX+Y)/(a+1)$, with $a=0.5$. 
In addition to this $A_{\rm eff}$ weighting, the second contribution comes from the system temperature. $T_{\rm sys}$ can vary in both the sky noise and the antenna (instrumental) noise components; the spillover term, which includes the ground-noise contribution, can also play a role in the observed response. Two test observations were performed at the SE607 station, each at a fixed $\rm EL = 60^\circ$ but either $15^\circ$ ahead or behind of the local Southern meridian in azimuth which corresponds to a variation of +1h and -1h in Local Sidereal Time (LST). The pointing direction is fixed at the two selected positions, while the sky drifted through the beam due to Earth's rotation. The noise component can then be measured. This shows a non-flat noise spectrum with respect to LST as shown in Figure~\ref{fig:tobia}. This can result in either higher or lower noise levels, and therefore S/N, depending on the observing time. Furthermore, as reported by \cite{Creaner_2019}, the presence of bright sources in the sidelobes affects the overall performance \citep[which can be degraded further by improper beam modelling as in the case of malfunctioning antennas, see][]{Brackenhoff_2025}, not only in azimuth but across different frequencies. 
A plausible explanation is that the asymmetric response arise from a combined effect of $\rm (a*X+b*Y)/T_{sys}$, where the $\rm a$ and $\rm b$ factors account for differences in the contributions from the two channels and $\rm T_{sys}$ embeds the noise factors from the sky and the antennas.

\section{Conclusions}\label{conclusions}

In this paper, we presented the first comprehensive analysis of the sensitivity of the HBAs of an international LOFAR station. The sensitivity of a radio telescope can be parametrized by determining the S/N of strong radio sources.

\begin{itemize}
    \item We use a sample of 11 bright, well known, pulsars, regularly distributed at equally spaced sky declinations. 
    %These are bright and well known pulsars. 
    They were monitored for several hours using the IE613 station, using the HBA in the frequency range $110-190$~MHz. The integrated pulse profiles were cleaned in time and frequency and the S/N retrieved through \texttt{pdmp}.  
    \item The response of the IE613 station displayed in Fig. ~\ref{Fig.synoptic_plot}, Fig. ~\ref{Fig.B1508+55_S/Nvsel_15min} and ~\ref{Fig.B1508+55_S/Nvsel_10MHz} shows a clear pattern in altitude and azimuth: over the entire bandwidth, improved S/N values are obtained near the zenith, but along the track the rising side has an overall better response than the setting one, with a discrepancy that reaches $45\%$.
    \item Other LOFAR stations were employed to investigate the single-station response pattern: dedicated observations were performed at SE607 for PSR~B0329$+$54, while observations of PSR~B0329$+$54 and PSR~B1508$+$55 at PL612 from \citet{Blaszkiewicz_2018} were used for comparison. The same azimuthal asymmetric and elevation-dependent pattern was observed at these stations.
    \item The Hamaker beam model and the Jones elements were evaluated to assess the hysteresis pattern and what may cause it. The observed asymmetry in azimuth is more than one can explain/model from this. Possible improper weighting of the X and Y components could explain such a discrepancy in S/N values. Noise levels can vary due to contamination of bright sources in the sidelobes which could be at the level observed, but would not have the observed azimuth dependence. 
    \item In the absence of a `fix' for the asymmetry, from a telescope operations point of view and in considering observing efficiencies with international LOFAR stations it then makes sense to observe sources as they rise, rather than when they set. With sensitivity that is tens of percent better on the `early' side, and the slow scaling of S/N with observing time, i.e. $\propto \sqrt{T_{\rm obs}}$, the relative time savings can be significant. 
\end{itemize}   

This study highlights that the capability of a radio telescope such as LOFAR strongly depends on the station beam model (which is narrower at the zenith), the geometrical projection of the collecting area (which is maximum at the zenith and diminishes as the pointing direction approaches the horizon), the frequency of the observation and the system noise (the radio sky is `louder' at lower frequencies). These effects can bias the data interpretation, as for flux estimation and, on larger scale, spectral-index measurements.

Investigating the sensitivity of LOFAR international stations, and identifying issues at the levels of tens of percent in response variation, leads us to consider the broader question of the response of upcoming and future low-frequency aperture-array telescopes, like the upgraded LOFAR2.0 and SKA-Low~\citep{LOFAR2}. These systems will work in tandem throughout the 2030s and beyond, LOFAR 2.0 in the North and SKA-Low in the South. SKA-Low will have better sensitivity, LOFAR 2.0 will have better angular resolution~\citep{Morabito_2025}. For such complex systems it can be quite a challenge to get everything working together --- we suggest tests such as we have presented here as ideal for early stage validation of these new systems. These can be performed even when the entire system is far from complete, e.g. in the case of SKA-Low as early as the second array assembly planned for early 2027.

%The LOFAR telescope is currently undergoing a system upgrade to LOFAR 2.0, which involves upgrades of the receiver electronics, transient buffer boards and more besides to enable parallel observations with the LBAs and HBAs; new automated observing and data processing pipeline software is also being rolled out~\citep{LOFAR2}. LOFAR 2.0 will operate as a Northern hemisphere counterpart to SKA-Low: SKA-Low will have better sensitivity, LOFAR will have have better angular resolution~\citep{Morabito_2025}. 

%On the other hand, 
SKA-Low telescope will soon be operating in the frequency range [$50-350$] MHz. With $131,072$ dual-polarised antennas, in form of log-periodic dipoles, spread over 512 stations, it will be the largest tied-array interferometer ever built and will bring unprecedented sensitivity over the $300$ MHz available bandwidth, over two orders of magnitudes higher than the current offered by LOFAR. As an aperture array, it's capability depends on frequency and pointing direction, as well as LST, with the sky noise being a major contributor to the antenna temperature \citep{Sokolowski_2022}. Given the orthogonal configuration of the dipoles, the two polarization channels will inevitably display different responses. However, `parallel transport aligned linear dipole' stations (e.g. LOFAR LBA) are not equivalent to stations with digitally beam-formed aligned dipoles (e.g. SKA-Low). In the latter, antennas are physically randomly orientated and aligned digitally. This approach mitigates the systematic polarization errors and improves calibration performance. Examining how the response of the antennas varies in frequency and pointing direction, addressing beam model errors and station-to-station variations, would reveal possible direction-dependent gain fluctuations. Unmodelled sky noise fluctuations and spatial beam errors will additionally constitute a source for instrumental-driven impurities in the signal, affecting high precision studies \citep[as for pulsar timing e.g.][]{vanderWateren_2023}, precise polarimetry and transient searches. This work highlights a potential limitation in radio interferometric measurements of aperture arrays with fixed dipoles. Similar tests could serve as a valuable diagnostic tool for identifying instrumental limitations and guiding future calibration improvements. Indeed, tests using pulsar observations as the one presented in this manuscript will raise potential issues and guide further investigation on their causes.

\section*{Acknowledgments}
LV acknowledges the support of a Lindsay Scholarship, funded by Armagh Observatory \& Planetarium, Dublin Institute for Advanced Studies and the School of Physics at Trinity College Dublin. OAJ acknowledges the support of the School of Physics at Trinity College Dublin and Breakthrough Listen which is managed by the Breakthrough Prize Foundation. The Rosse Observatory is operated by Trinity College Dublin; we acknowledge support from a Higher Education Research Equipment Grant (PI: Keane). I-LOFAR infrastructure has benefited from funding from Science Foundation Ireland, a predecessor of Taighde Éireann — Research Ireland. We acknowledge support from Onsala Space Observatory for the provisioning of its facilities and observational support. The Onsala Space Observatory national research infrastructure is funded through Swedish Research Council grant No 2017-00648.\\
\\
\textit{Author contributions.}\\
The work presented here uses data collected using the Irish LOFAR telescope ($\sim$170 hours) and the Swedish LOFAR telescope (9 hours) when in long-term standalone mode during the LOFAR 2.0 upgrade. The work benefited from, and built upon, a large pre-existing software base developed by many authors over many years, in particular software developers in the pulsar science community and at ASTRON. 

Contributions by individuals specific to the work presented in this paper, as opposed to the ongoing effort to operate the telescopes, are listed below in the CRedIT\footnote{\url{https://credit.niso.org/}} format.\\

\textit{CRediT statement:}\\
LV: Methodology, Software, Validation, Formal Analysis, Investigation, Data Curation, Writing --- Original Draft, Writing --- Review \& Editing, Visualization. \\
SCS: Software, Investigation, Data Curation, Writing --- Review \& Editing. \\
OAJ: Software, Validation, Investigation, Data Curation, Writing --- Review \& Editing.  \\
EFK: Conceptualization, Methodology, Investigation, Resources, Writing --- Original Draft, Writing --- Review \& Editing, Supervision, Project Administration, Funding Acquisition. \\
DJMcK: Conceptualization, Software, Resources, Writing --- Review \& Editing. \\
TDC: Software, Investigation, Resources. \\
GR: Writing --- Review \& Editing, Supervision, Funding Acquisition.\\
OC: Conceptualization, Software, Writing --- Review \& Editing. \\
PTG: Supervision, Funding Acquisition\\
JMcC: Software, Resources, Project Administration\\

\bibliographystyle{apsrev4-1}
\bibliography{ie613_paper}

\clearpage
\mbox{}
\clearpage
\begin{appendix}

\section{Appendix A: Plot of S/N for all pulsars in the sample for I-LOFAR station}\label{app:B}

\begin{figure}[h!]
   \centering
   \includegraphics[width=13cm]{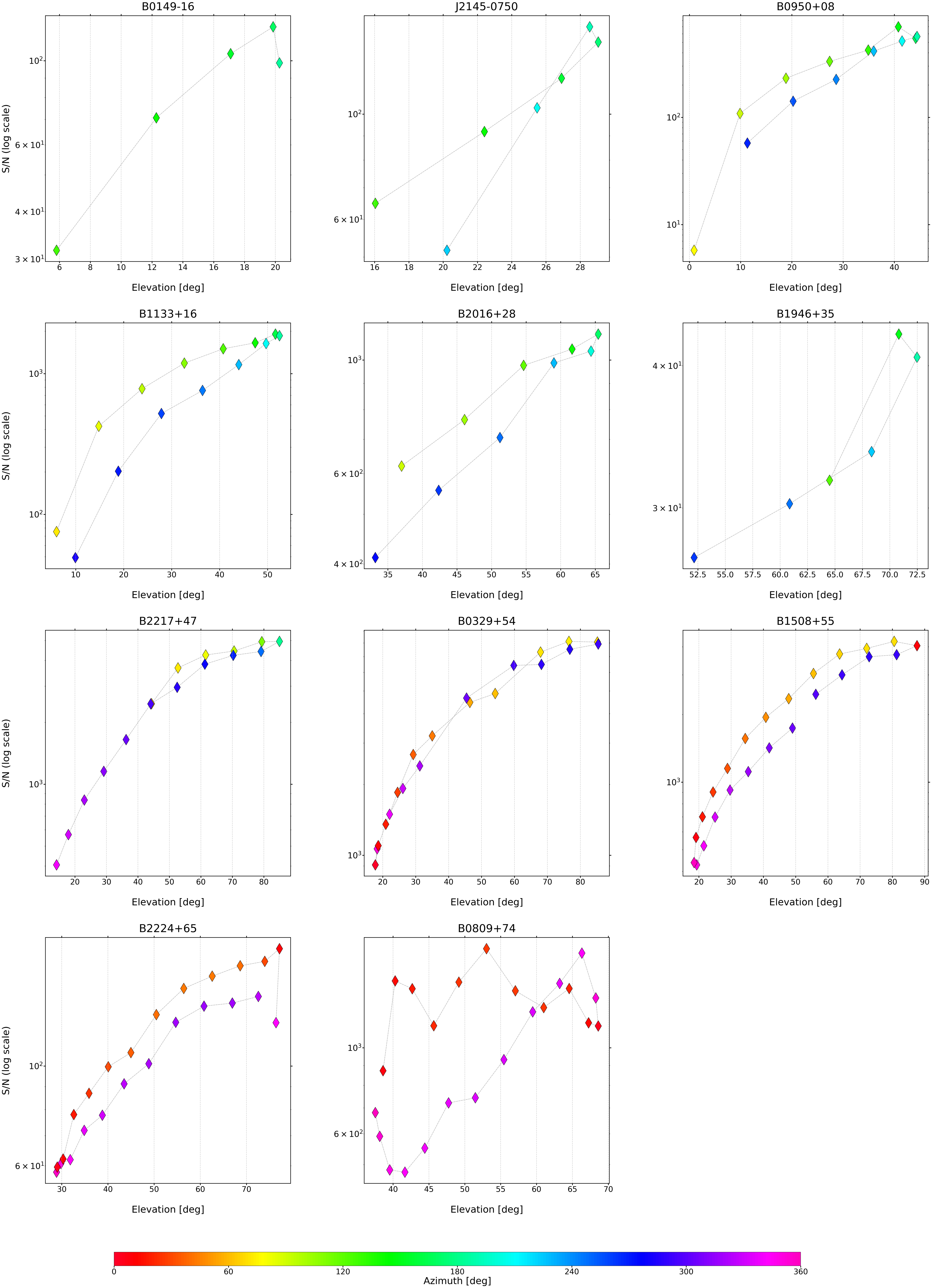}
   \caption{S/N vs elevation in semi-log scale. Each pulsar is shown in a single subplot, with the S/N on the y-axis (log scale) and the elevation on the x-axis. For each subplot, the points represent the S/N of the RFI-cleaned integrated pulse profiles over 1-hour and for the full frequency range (reduced to $112-190$~MHz) and colour-coded by azimuth.}
     \label{fig:all_snr_azimuth}
\end{figure}
\newpage
\section{Appendix B: Retrieved S/N from post-processed data }\label{app:A}
\begin{table}[h!]
\centering
\caption{Summary of the pulsars data, with pulsar name in B1950 coordinate system \textbf{(1)}, elevation in degrees \textbf{(2)}, azimuth in degrees \textbf{(3)} and relative S/N \textbf{(4).}}\label{table:appA}        
    \begin{tabular}{l l l c | l l l c}
        \toprule\hline
        PSR & EL[deg] & Az [deg] & S/N & PSR & EL[deg] & Az [deg] & S/N \\
        \hline & \\[-0.01ex]
        \multirow[t]{7}{*}{\underline{B0149-16}}
                & 5.8 & 127.7 & 31.65 &  \multirow[t]{7}{*}{\underline{J2145-0750}} & 16.0 & 128.0 & 64.88 \\
                & 12.3 & 140.7 & 70.72 & & 22.4 & 142.3 & 91.88 \\
                & 17.1 & 154.8 & 104.36 &  & 26.9 & 157.9 & 119.01 \\
                & 19.9 & 169.7 & 122.87 &  & 29.1 & 174.7 & 141.83 \\
                & 20.3 & 185.1 & 98.69 & & 28.6 & 191.7 & 152.73 \\
                & & & & & 25.5 & 208.1 & 103.11 \\
                & & & & & 20.2 & 223.2 & 51.67 \\ 
        \hline & \\[-0.01ex]
        %PSR & (EL,AZ) [deg,deg] & S/N & PSR & (EL,AZ) [deg,deg] & S/N \\
        \multirow[t]{13}{*}{\underline{B0950+08}}
                & 0.9 & 78.2 & 5.76 & \multirow[t]{13}{*}{\underline{B1133+16}} & 6.0 & 71.7 & 75.37 \\
                & 9.9 & 90.1 & 108.6 && 14.8 & 83.4 & 422.91 \\
                & 18.8 & 102.4 & 231.56 && 23.8 & 95.4 & 780.77 \\
                & 27.4 & 115.8 & 332.15 && 32.6 & 108.5 & 1186.82 \\
                & 34.9 & 130.8 & 423.62 && 40.8 & 123.5 & 1501.34 \\
                & 40.8 & 148.2 & 698.26 && 47.4 & 141.5 & 1655.78 \\
                & 44.2 & 167.9 & 545.52 && 51.6 & 162.9 & 1908.43 \\
                & 44.4 & 188.8 & 566.95 && 52.5 & 186.5 & 1859.24 \\
                & 41.5 & 208.8 & 514.41 && 49.7 & 209.2 & 1636.96 \\
                & 36.0 & 226.6 & 414.37 && 44.0 & 228.8 & 1158.49 \\
                & 28.7 & 242.0 & 225.77 && 36.4 & 245.0 & 759.18 \\
                & 20.2 & 255.5 & 141.32 && 27.9 & 258.8 & 520.48 \\
                & 11.3 & 268.0 & 57.54 && 18.9 & 271.2 & 202.6 \\
                &&&&& 9.9 & 283.0 & 49.31 \\ 
        \hline & \\[-0.01ex]
        \multirow[t]{10}{*}{\underline{B2016+28}}
                & 37.0 & 90.0 & 620.98 & \multirow[t]{6}{*}{\underline{B1946+35}} & 64.5 & 122.2 & 31.73 \\
                & 46.1 & 103.2 & 765.1 & & 70.8 & 150.2 & 42.64 \\
                & 54.6 & 119.5 & 976.29 & & 72.5 & 189.3 & 40.69 \\
                & 61.6 & 141.4 & 1050.33 & & 68.3 & 224.1 & 33.62 \\
                & 65.4 & 170.5 & 1123.55 & & 60.9 & 247.0 & 30.26 \\
                & 64.4 & 202.4 & 1040.82 & & 52.2 & 262.6 & 27.14 \\
                & 59.0 & 228.5 & 987.08 &&&&\\
                & 51.2 & 247.8 & 705.86 &&&&\\
                & 42.4 & 262.5 & 556.84 &&&&\\
                & 33.2 & 274.9 & 411.66 &&&& \\
        \hline & \\[-0.01ex]
        \multirow[t]{16}{*}{\underline{B2217+47}}
                & 44.3 & 64.5 & 2463.55 & \multirow[t]{28}{*}{\underline{B0329+54}} & 18.3 & 352.0 & 1064.59 \\
                & 52.7 & 72.8 & 3679.18 && 17.8 & 1.1 & 909.71 \\
                & 61.5 & 81.8 & 4252.43 && 18.7 & 10.2 & 1098.25 \\
                & 70.5 & 92.9 & 4445.3 & & 20.9 & 19.1 & 1354.55 \\
                & 79.3 & 112.3 & 4934.48 & & 24.5 & 27.7 & 1851.15 \\
                & 84.9 & 181.7 & 4951.83 && 29.3 & 35.8 & 2682.65 \\
                & 79.1 & 248.6 & 4422.61 && 35.0 & 43.5 & 3220.97 \\
                & 70.3 & 267.5 & 4235.88 && 46.5 & 55.2 & 4470.41 \\
                & 61.3 & 278.5 & 3836.63 && 54.1 & 61.5 & 4878.0 \\
                & 52.5 & 287.4 & 2960.56 && 67.9 & 70.6 & 7321.95 \\
                & 44.1 & 295.7 & 2458.77 && 76.5 & 74.2 & 8131.03 \\
                & 36.2 & 304.0 & 1648.85 && 85.2 & 67.9 & 8078.47 \\
                & 29.2 & 312.5 & 1155.39 && 85.4 & 293.0 & 7920.51 \\
                & 23.0 & 321.3 & 837.84 & & 76.8 & 285.8 & 7525.16 \\
                & 17.9 & 330.5 & 568.9 && 68.2 & 289.2 & 6487.54  \\
                &&&&& 59.8 & 294.4 & 6423.67 \\ 
                &&&&& 45.5 & 305.7 & 4665.34 \\
                &&&&& 31.3 & 321.3 & 2399.57 \\
                &&&&& 26.1 & 329.3 & 1924.03 \\
                &&&&& 22.1 & 337.8 & 1494.13 \\
                &&&&& 19.3 & 346.6 & 1074.96 \\
                &&&& & 17.9 & 355.6 & 959.58 \\
                &&&& & 17.9 & 4.7 & 935.37 \\
                &&&& & 19.4 & 13.8 & 1185.88 \\
                &&&&& 22.2 & 22.5 & 1795.75 \\
                &&&&& 26.3 & 31.0 & 2114.08 \\
                &&&&& 31.5 & 38.9 & 2895.92 \\
                &&&&& 37.6 & 46.4 & 3149.01 \\
        \bottomrule
    \end{tabular}
\end{table}

\begin{table}[h]
    \centering
    \begin{tabular}{l l l c | l l l c}
    \toprule\hline
        PSR & EL[deg] & Az [deg] & S/N & PSR & EL[deg] & Az [deg] & S/N \\
        \hline & \\[-0.01ex]
        \multirow[t]{25}{*}{\underline{B1508+55}}  
                & 49.0 & 303.9 & 1740.06 & \multirow[t]{24}{*}{\underline{B2224+65}}& 72.6 & 325.7 & 142.74 \\
                & 41.8 & 310.4 & 1420.48 && 66.9 & 318.4 & 137.99 \\
                & 35.3 & 317.5 & 1113.33 && 60.8 & 316.8 & 135.85 \\
                & 29.7 & 325.0 & 921.71 && 54.7 & 318.3 & 125.04 \\
                & 25.0 & 333.1 & 698.53 && 48.9 & 321.5 & 101.18 \\
                & 21.5 & 341.5 & 520.63 && 43.5 & 326.0 & 91.28 \\
                & 19.3 & 350.3 & 428.08 && 38.8 & 331.3 & 77.73 \\
                & 18.5 & 359.3 & 438.25 && 34.9 & 337.3 & 71.94 \\
                & 19.1 & 8.3 & 566.95 & & 31.9 & 343.7 & 61.91 \\
                & 21.1 & 17.1 & 700.43 && 29.8 & 350.5 & 60.71 \\
                & 24.4 & 25.6 & 903.36 && 28.9 & 357.5 & 58.09 \\
                & 28.9 & 33.7 & 1150.42 && 29.0 & 4.5 & 59.62 \\
                & 34.4 & 41.4 & 1563.72 && 30.3 & 11.5 & 62.13 \\
                & 40.8 & 48.5 & 1942.81 && 32.6 & 18.2 & 78.0 \\
                & 47.9 & 55.1 & 2352.13 & & 35.9 & 24.5 & 86.96 \\
                & 55.5 & 61.1 & 3043.62 && 40.1 & 30.3 & 99.61 \\
                & 63.6 & 66.4 & 3718.18 && 45.0 & 35.4 & 107.09 \\
                & 72.0 & 70.2 & 3937.6 && 50.5 & 39.5 & 130.07 \\
                & 80.5 & 69.4 & 4231.42 && 56.4 & 42.4 & 148.7 \\
                & 87.6 & 9.4 & 4048.59 && 62.6 & 43.1 & 158.36 \\
                & 81.3 & 291.3 & 3692.11 & & 68.6 & 40.3 & 167.0 \\
                & 72.8 & 289.6 & 3615.52 & & 74.0 & 30.4 & 170.82 \\
                & 64.4 & 293.2 & 3002.11 && 77.1 & 9.0 & 182.27 \\
                & 56.2 & 298.4 & 2460.05 & & 76.4 & 342.0 & 124.82 \\
        \hline & \\[-0.01ex]
         \multirow[t]{23}{*}{\underline{B0809+74}} 
                & 38.6 & 8.4 & 872.57 &&&\\
                & 40.3 & 13.1 & 1486.67 &&&\\
                & 42.7 & 17.5 & 1419.41 &&&\\
                & 45.7 & 21.2 & 1139.42 &&&\\
                & 49.2 & 24.2 & 1475.66 &&&\\
                & 53.0 & 26.1 & 1799.94 &&&\\
                & 57.0 & 26.5 & 1402.05 &&&\\
                & 61.0 & 25.0 & 1269.14 &&&\\
                & 64.6 & 20.8 & 1421.0 &&&\\
                & 67.3 & 13.5 & 1159.27 &&&\\
                & 68.6 & 3.3 & 1137.84 &&&\\
                & 68.3 & 352.4 & 1343.63 &&&\\
                & 66.3 & 343.2 & 1753.17 &&&\\
                & 63.2 & 337.2 & 1464.58 &&&\\
                & 59.5 & 334.1 & 1237.26 &&&\\
                & 55.5 & 333.4 & 931.94 &&&\\
                & 51.5 & 334.5 & 743.26 &&&\\
                & 47.7 & 336.9 & 720.95 &&&\\
                & 44.4 & 340.2 & 551.27 &&&\\
                & 41.7 & 344.2 & 478.06 &&&\\
                & 39.5 & 348.7 & 484.47 &&&\\
                & 38.1 & 353.6 & 591.25 &&&\\
                & 37.5 & 358.6 & 680.58 &&&\\
            \bottomrule
    \end{tabular}
\end{table}

\end{appendix}

\end{document}